\documentclass[sigconf, nonacm]{acmart}
\usepackage{xcolor,listings}
\usepackage{algorithm} 
\usepackage{booktabs}
\usepackage[noend]{algpseudocode}
\usepackage{pdfpages} 
\usepackage{graphicx}
\usepackage{tabularx}
\usepackage{subcaption}
\usepackage[T1]{fontenc}
\usepackage{upquote}
\newcommand\vldbdoi{XX.XX/XXX.XX}
\newcommand\vldbpages{XXX-XXX}
\newcommand\vldbvolume{14}
\newcommand\vldbissue{1}
\newcommand\vldbyear{2020}
\newcommand\vldbauthors{\authors}
\newcommand\vldbtitle{\shorttitle} 
\begin{document}
\title{Revisiting Runtime Dynamic Optimization for Join Queries in Big Data Management Systems}

\author{Christina Pavlopoulou}
\affiliation{%
  \institution{University of California, Riverside}
}
\email{cpavl001@ucr.edu}

\author{Michael J. Carey}
\affiliation{%
  \institution{University of California, Irvine}
}
\email{mjcarey@uci.edu}

\author{Vassilis J. Tsotras}
\affiliation{%
  \institution{University of California, Riverside}
}
\email{tsotras@cs.ucr.edu}

\begin{abstract}

Query Optimization remains an open problem for Big Data Management Systems.
Traditional optimizers are cost-based and use statistical estimates of intermediate result cardinalities to assign costs and pick the best plan. However, such estimates tend to become less accurate because of filtering conditions caused either from undetected correlations between multiple predicates local to a single dataset, predicates with query parameters, or predicates involving user-defined functions (UDFs). Consequently, traditional query optimizers tend to ignore or miscalculate those settings, thus leading to suboptimal execution plans. Given the volume of today's data, a suboptimal plan can quickly become very inefficient. 

In this work, we revisit the old idea of \textit{runtime dynamic optimization} and adapt it to 
a shared-nothing distributed database system, AsterixDB. 
The optimization runs in stages (re-optimization points), starting by first executing 
all predicates local to a single dataset.
The intermediate result created from each stage is used to re-optimize the remaining query. 
This re-optimization approach avoids inaccurate intermediate result cardinality estimations, thus leading to much better execution plans.
While it introduces the overhead for materializing these intermediate results, 
our experiments show that this overhead is relatively small and it is an acceptable price to pay given the optimization benefits.  
In fact, our experimental evaluation shows that runtime dynamic optimization leads to much better execution plans as compared to the current default AsterixDB plans as well as to plans produced by static cost-based optimization (i.e. based on the initial dataset statistics) and other state-of-the-art approaches.

\end{abstract}

\maketitle

\begingroup\small\noindent\raggedright\textbf{PVLDB Reference Format:}\\
\vldbauthors. \vldbtitle. PVLDB, \vldbvolume(\vldbissue): \vldbpages, \vldbyear.\\
\href{https://doi.org/\vldbdoi}{doi:\vldbdoi}
\endgroup
\begingroup
\renewcommand\thefootnote{}\footnote{\noindent
This work is licensed under the Creative Commons BY-NC-ND 4.0 International License. Visit \url{https://creativecommons.org/licenses/by-nc-nd/4.0/} to view a copy of this license. For any use beyond those covered by this license, obtain permission by emailing \href{mailto:info@vldb.org}{info@vldb.org}. Copyright is held by the owner/author(s). Publication rights licensed to the VLDB Endowment. \\
\raggedright Proceedings of the VLDB Endowment, Vol. \vldbvolume, No. \vldbissue\ %
ISSN 2150-8097. \\
\href{https://doi.org/\vldbdoi}{doi:\vldbdoi} \\
}\addtocounter{footnote}{-1}\endgroup

\section{Introduction}

Query optimization is a core component in traditional database systems, as it facilitates the order of execution decisions between query operators along with each operator's physical implementation algorithm. One of the most demanding operators is the Join, as it can be implemented in many different ways depending on the sizes of its inputs and outputs. To tackle the join optimization problem, 
two different approaches have been introduced.

The first approach (introduced in System R \cite{r}) is cost-based query optimization; it performs an exhaustive search (through dynamic programming) among all different join orderings 
until the one with the smallest cost is found and eventually executed in a pipelined mode. 
The second approach (introduced around the same time in INGRES \cite{ingres}) uses instead
a runtime dynamic query optimization method (later known as 
Adaptive Query Processing (AQP)), where the original query is decomposed into single-variable (i.e., single dataset) subqueries which are executed separately. 
This decomposition takes place in the following ways: (1)
breaking off components of the query which are
joined to it by a single variable, (2) substituting for one of the variables
a tuple-at-a-time (to perform the join operation).
Each subquery result is stored as a 
new relation that is then considered by the optimizer so as to optimize the remaining query. 
The choice of the ``next" subquery to be executed is based on the cardinality of the participating datasets. 

The INGRES approach was a greedy cardinality-based technique, with runtime overhead due to creating indexed (for joins) intermediate results, and the more comprehensive, cost-based, compile-time approach of System-R became the field's preferred approach \cite{vertica, orca, graefe1995cascades, volcano} for many years.
To assign a cost for each plan (and thus find the best join ordering and implementation algorithms among the search space) the cost-based approach depends heavily on statistical information. 
The accuracy of such statistics is greatly affected by the existence of
 multiple selection predicates (on a single dataset), \textit{complex} selection predicates (i.e., with parameterized values or UDFs) and join conditions that are not based on key-foreign key relationships. 
In such cases, statistics can be very misleading, resulting in inaccurate join result estimations. As the number of joins increases,
 the error can get worse as it gets propagated to future join stages \cite{ioannidis}. These issues are exacerbated in today's big data management systems (BDMS) by the sheer volume of data.

In this work, we revisit the runtime dynamic optimization introduced by INGRES \cite{ingres} and adapt it (with modifications) to a shared-nothing distributed BDMS, namely, AsterixDB.
With the increase in the volume of data, even small errors in the join order can generate \textit{very expensive} execution plans. 
A characteristic of the original dynamic optimization approach is that the choice of the "next" subquery to be executed is based only on dataset cardinality. 
However, the alternative cost-based optimization approach has shown that, for better join result estimation, one needs better statistics. 
Thus, we take advantage here of the materialization stages to collect all needed statistics. 
This combination of re-optimization and statistics collection leads to superior execution plans.

Specifically, when a query is executed, all predicates local to a table are pushed down and they are executed first to gather updated accurate statistics. 
The intermediate results along with the updated statistics are fed back to the optimizer to choose the cheapest initial join to be executed. 
This process is repeated until only two joins are left in the query. 
We integrated our techniques in AsterixDB \cite{asterix, asterixdb} which, like many relational database systems, is optimized for executing queries in a pipelined manner. 
Although with our modified dynamic optimization approach the query execution goes through blocking re-optimization points, this extra overhead is relatively minimal and is thus worthwhile since very expensive query plans are avoided.  

Various works have been proposed in literature that use dynamic optimization techniques to alleviate the problems introduced by static cost-based optimization \cite{ss, aqp, mid, eddies, leo}. In this context, new statistics are estimated after mid-query execution (with information gathered from intermediate results) and they are used to re-calibrate the query plan. This is similar to our approach; however, such works tend to ignore information coming from correlated selectivities, predicates with parameterized values and UDFs. Instead, by executing the local predicates first, we gain accurate cardinality estimations early that lead to improved query performance (despite the overhead of materializing those filters).
Dynamic optimization has also been introduced in multi-node environments \cite{rios, pilot, rope}. These works
either introduce unnecessary additional overheads by running extra queries to acquire statistical data for the datasets \cite{pilot} or they need to re-partition data because of lazily picking an inaccurate  initial query plan \cite{rios}. Optimus \cite{optimus} also uses runtime dynamic optimization, but it does not consider queries with multiple joins. Re-optimization points are used in \cite{rope} in a different way, as a place where an execution plan can be stopped if its execution is not as expected. 

As we show in the experimental evaluation, for a variety of workloads, our modified runtime dynamic optimization will generate query plans that are better than even the best plans formed by (i) a user-specified order of the datasets in the FROM clause of a submitted query, or (ii)  traditional static cost-based optimizers. In particular, our methods prevent the execution of expensive plans and promote more efficient ones. Re-optimizing the query in the middle of its execution and not focusing only on the initial plan can be very beneficial, as in many cases, the first (static) plan is changed dramatically by our optimizer. 

In summary, this paper makes the following contributions: 
\begin{itemize}
    \vspace{-1em}\item We adapt an INGRES-like dynamic optimization scheme in a shared-nothing BDMS (AsterixDB). 
    This includes a predicate pre-processing step that accurately estimates initial selectivities by executing all predicates local to a dataset early on. 
    We insert multiple re-optimization points during query execution to receive feedback (updated statistics for join results) and refine the remaining query execution plan. 
    At each stage (i.e. re-optimization point), we only consider the next cheapest join, thus avoiding forming the whole plan and searching among all the possible join ordering variations.
    
    \item We assess the proposed dynamic optimization approach via detailed experiments that showcase its superiority against traditional optimizers. We also evaluate the overhead introduced by the multiple re-optimization points and the materialization of intermediate results.
\end{itemize}

The rest of the paper is organized as follows: Section \ref{Related Work} discusses existing work on runtime dynamic optimization, while Sections \ref{AsterixDB}  and \ref{col} outline the architecture of AsterixDB and the statistics collection framework respectively. Section \ref{Solution} describes the details of the dynamic optimization approach including the use of statistics, while Section \ref{Integration} showcases how the approach has been integrated into the current version of AsterixDB. The experimental evaluation appears in Section \ref{Experiments}. Section \ref{Conclusion} concludes the paper and presents directions for future research.

\vspace{-2em}
\section{Related Work}
\label{Related Work}

Traditional query optimization focuses on cost models derived from statistics on base datasets (cost-based optimization) as introduced in System R \cite{r}. Typically, there are two steps in this process: first, there is a rewrite phase that transforms the specified query into a collection of alternate plans (created by applying a collection of rules), and second, cost models based on cardinality estimation are used to pick the plan with the least cost \cite{graefe1995cascades, graefe1987exodus, memsql}. A cost-based optimization approach adapted for parallel shared-nothing architectures is described in \cite{spread}; here the master node sends the query to all worker nodes along with statistics. Then, each worker decides the best plan based on its restrictions and sends its decision to the master. Finally, the master decides the globally optimal plan. This way, all the nodes in the cluster are working in parallel to find the best plan, each node working with a smaller set of plans. Our work also considers the shared-nothing environment, however, we concentrate on runtime dynamic optimization.

Runtime dynamic optimization was introduced in INGRES \cite{ingres}, where a query is decomposed into single-variable queries (one dataset in the FROM clause) which are executed separately. Based on the updated intermediate data cardinalities, the next best query is chosen for execution. In our work, we wanted to revisit this approach and see whether big data processing systems can benefit from it.
Hence we execute part of the query to obtain statistics from the intermediate results and refine the remaining query. Opposite to INGRES, we do not depend only on cardinalities to build our cost model, but we collect more information regarding base and intermediate data based on statistics. Since INGRES, there have been various works using runtime dynamic optimization in a single-server context. Specifically, LEO \cite{leo} calibrates the original statistics according to the feedback acquired from historical queries and uses them to optimize future queries. In Eddies \cite{eddies} the selectivity of each query operator is calculated while records are being processed. Eventually, the more selective operators are prioritized in the evaluation order.

Dynamic optimization is more challenging in a 
shared-nothing 
environment, as data is kept and processed across multiple nodes.
Optimus \cite{optimus} leverages runtime statistics to rewrite its execution plans. Although it performs a number of optimizations, 
it does not address multi-way joins, which as \cite{optimus}
points out, can be “tricky” because the data may need to
be partitioned in multiple ways.

RoPE \cite{rope} leverages historical statistics from prior plan executions in order to tune future executions, e.g. the number of reduce tasks to schedule, choosing appropriate operations, including order. Follow-up work \cite{bruno} extends the RoPE design to support general query workloads in Scope \cite{scope}. Their strategy generates a (complete) initial query plan from historical statistics, and it collects fresh statistics (specifically, partitioned histograms) during execution that can be used to make optimized adjustments to the remaining operators in the plan. However, in order not to throw away work, reoptimization takes place after a certain threshold and the initial plan is configured only based on the base datasets, which can potentially lead to suboptimal plans. 
In contrast, in our approach we block the query after each join stage has been completed and we use the result to optimize the subsequent stages; hence no join work is wasted. Furthermore, we estimate the selectivity of predicates by pushing down their execution; hence we avoid initial possibly misleading calculations. Nevertheless, learning from past query executions is an orthogonal approach that could be used to further optimize our approach and it is part of our future work.

Another approach belonging to the runtime dynamic optimization category uses \textit{pilot runs}, as introduced in 
\cite{pilot}. In an effort to alleviate the need for historical statistics, pilot runs of the query are used on sample data. 
There are two main differences between this approach and our work. First, statistics obtained by pilot runs are not very accurate for joins that do not have a primary/foreign key condition as sampling can be skewed under those settings. 
In contrast, our work gathers statistics on the base datasets which leads to more accurate join result estimations for those joins. Secondly, in our work we exploit AsterixDB's LSM ingestion process to get initial statistics for base datasets along with materialization of intermediate results 
to get more accurate estimations - thereby we avoid the extra overhead of pilot runs.

Finally, RIOS \cite{rios} is another system that promotes runtime incremental optimization. In contrast to Optimus, 
RIOS assumes that the potential re-partitioning overhead is amortized by the efficiency of their approach. Particularly, statistics are collected during a pre-partitioning stage in which all the datasets participating in the query are partitioned according to an initial lazy plan formed based on raw byte size.  
However, if later statistics (collected during the pre-partitioning stage) indicate that this is not the correct plan, RIOS re-partitions the data. This is done if and only if the difference between the lazy plan and the better one is larger than a certain threshold. 
In that case, the remaining query is optimized according to the feedback acquired by intermediate results. 
In contrast to RIOS, our method alleviates the need for potential expensive re-partitioning since accurate statistics are collected before the query is processed by the optimizer. That way, we can pick the right join order from the beginning and thereby the right partitioning scheme. Hence, we avoid the overhead of faulty partitioning, which for large volumes can be very significant.

\section{AsterixDB Background}
\label{AsterixDB}

\begin{figure}[h!]
\includegraphics[width=0.5\columnwidth, height=4cm]{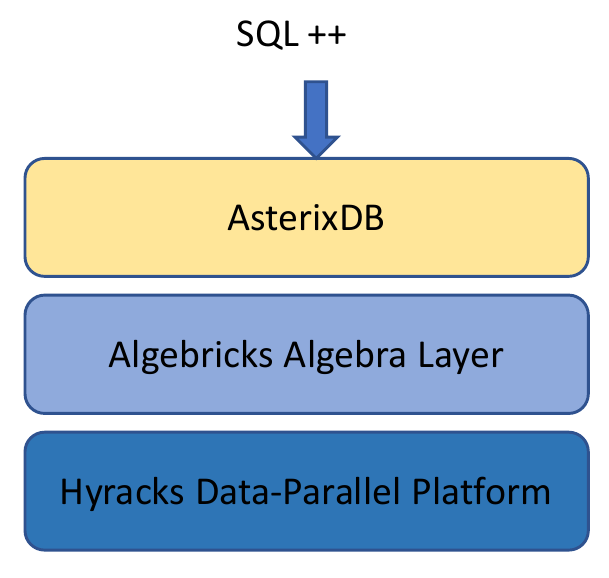}
\caption{AsterixDB Architecture}
\label{architecture}
\end{figure}

Apache AsterixDB is a parallel, shared-nothing platform that provides the ability to ingest, store, index, query, and analyze mass quantities of semistructured data. 
As shown in Figure \ref{architecture}, to process a submitted query, AsterixDB compiles it into an Algebricks \cite{algebricks} program also known as the logical plan. This plan is then optimized via rewrite rules that reorder the Algebricks operators and introduce partitioned parallelism for scalable execution. After this (rule-based) optimization step, a code generation step translates the resulting physical query plan into a corresponding Hyracks Job \cite{hyracks} that will use the Hyracks engine to compute the requested query results. Finally, the runtime plan is distributed accross the system and executed locally on every slave of the cluster. 

Although all AsterixDB layers will participate in the integration of our work, the query optimizer, which is mainly in the Algebricks layer, will be our core focus. Currently, the AsterixDB optimizer takes into consideration many data properties, such as the data partitioning and ordering, and decides according to a set of heuristic rules (which are the core of Algebricks) how the query should be executed.
These heuristic rules are applied without any information gathered from statistics. For multi-join queries, the join order in AsterixDB currently depends on the order of the datasets in the \textit{FROM} clause of the query (i.e., datasets are picked in the order they appear in it). Generally, the compiler will produce right-deep joins; if the user wants to generate bushy-joins, it is feasible by grouping the datasets together using parentheses. However, in our experience this option can be complicated for naive users.


Another aspect in join query optimization is the choice of join algorithm. AsterixDB supports multiple algorithms like Hash, Broadcast and Nested Loop Join. Below, we describe the implementation of each algorithm in AsterixDB. 

\textbf{Hash Join:} Assuming the join’s input data is not partitioned in a useful way, the algorithm redistributes the data by hashing both inputs on the join key(s) – thereby ensuring that objects that should be joined will be routed to the same partition for processing – and then effects the join using dynamic hash join. In more detail, the “build” side of the join is first re-partitioned and fed over the network into the build step of a local hash join; each partition will then have some portion (perhaps all) of the to-be-joined build input data in memory, with the rest (if any) in overflow partitions on disk. The “probe” side of the join is then re-partitioned similarly, thus creating a pipelined parallel orchestration of a dynamic hash join. In the event that one of the inputs is already partitioned on the join key(s), e.g., because the join is a key/foreign key join, re-partitioning is skipped (unnecessary) for that input and communication is saved. 

\textbf{Broadcast Join:} This strategy employs a local dynamic hash join where one of the join inputs (ideally a small one) is broadcast – replicated, that is – to all partitions of the other input. The broadcast input is used as the build input to the join, and once the build phase is done the participating partitions can each probe their local portion of the other larger input in order to effect the join. 

\textbf{Indexed Nested Loop Join:} Here, one of the inputs is broadcast (replicated) to all of the partitions of the other input, which for this strategy must be a base dataset with an index on the join key(s); as broadcast objects arrive at each partition they are used to immediately probe the index of the other (called “inner”) dataset.



Currently, in AsterixDB, the hash join is picked by default unless there are query hints that make the optimizer pick one of the other two algorithms. However, when a broadcast join can be applied, joins can complete much faster as expensive shuffling of the large dataset is avoided.

\textbf{Optimizer Limitations:} The current rule-based optimizer in AsterixDB has several limitations: 
\begin{itemize}
   \item There is no selectivity estimation for predicates. Consequently, opportunities are missed for choosing the right join orders and join algorithms. Broadcast joins, in particular, will not be considered without a hint, even in the case when a dataset becomes small enough to fit in memory after the application of a selective ﬁlter.
    \item There is no cost-based join enumeration. Thus, a query's performance relies largely on the way it has been written by the user (i.e., the dataset ordering in the FROM clause).
\end{itemize}

Note that the above limitations are present in other existing large scale data platforms as well. We expect that the techniques presented in this work would also be beneﬁcial for those systems. 

\section{Statistics Collection}
\label{col}

At each re-optimization point, we collect statistical information about the base and intermediate datasets that will help the optimizer decide the best join order and join algorithm. These statistics are later used to estimate the actual join result size by using the following formula, as described in \cite{selinger}: 

\begin{equation}
    A \bowtie_k B = S(A) * S(B) / max(U(A.k), U(B.k))
    \label{join}
\end{equation}

\noindent where $S(x)$ is the size of dataset $x$ and $U(x.k)$ is the number of unique elements for attribute $k$ of dataset $x$. The size of a dataset is the number of qualified records in the dataset immediately before the join operation. If a dataset has local predicates, the traditional way to calculate result cardinality
is to multiply all the individual selectivities \cite{selinger}. However, as it will be described in section \ref{predicates}, we use a more effective approach for this calculation. 


\textbf{Statistics Types:} To measure the selectivity of a dataset for specific values, we use quantile sketches. Following the Greenwald-Khanna algorithm \cite{gk}, we extract quantiles which represent the right border of a bucket in an equi-height histogram. The buckets help us identify estimates for different ranges which are very useful in the case that filters exist in the base datasets.
To find the number of unique values needed for formula \ref{join}, we use Hyperloglog \cite{hll} sketches. The HLL algorithm can identify with great precision the unique elements in a stream of data. We collect these types of statistics for every field of a dataset that may participate in any query. It should be noted that the gathering of these two statistical types happens in parallel. 


\section{Runtime Dynamic Optimization}
\label{Solution}

The main focus of our dynamic optimization approach is to utilize the collected statistics from intermediate results in order to refine the plan on each subsequent stage of a multi join query. To achieve this aim, there are several stages that need to be considered. 


As described in Algorithm \ref{algo} lines 6-9, the first step is to identify all the datasets with predicates. If the number of predicates is more than one, or, there is at least one complex predicate (with a UDF or parameterized values), we execute them as described in Section \ref{predicates}. 
Afterwards, while the updated query execution starts as it would normally do, we introduce a loop which will complete only when there are only two joins left in the query. In that case, there is no reason to re-optimize the query as there is only one possible remaining join order. This loop can be summarized in the following steps: 
\begin{itemize}
    \item A query string, along with statistics, are given to the \textbf{Planner} (line 12) which is responsible for figuring out the next best join to be executed (the one that results in the least cardinality)  based on the initial or online statistics. As a result, the Planner does not need to form the complete plan, but only to find the cheapest next join for each iteration.
    \item The output plan is given as input to the \textbf{Job Construction} phase (line 14) which actually converts it to a job (i.e. creation of query operators along with their connections). This job is executed and the materialized results will be rewired as input whenever they are needed by subsequent join stages.
    \item Finally, if the remaining number of datasets is more than three, we return to the \textbf{Planner} phase with the new query as formatted in the \textbf{Query Reconstruction} phase (line 13); otherwise the result is returned. 
\end{itemize}

\begin{algorithm}
	\caption{Dynamic Optimization} 

	\begin{algorithmic}[1]
	\small
	\State $J\leftarrow$ joins participating in the original query
	\State $D\leftarrow$ collection of base datasets ($d$) in the query
	\State $Statistics\leftarrow$ quantile and hyperloglog sketches for each field of $D$ that is a join key
	\State $Q(\sigma,D,J) \leftarrow$ original query as submitted by user\Comment{$\sigma$ is the projection list}\\
	
	\For{$d$ in $D$}
	    \State $P\leftarrow$ set of selective predicates local to $d$
	    \If{$|P|$ > 1}
	       \State $D-\{d\}\bigcup$\Call{PushDownPredicates}{$d,P$}
	    \EndIf
	\EndFor \\
	
	\While{$|J|$ > 2} 
		\State $j\leftarrow$ \Call{Planner}{J, Statistics} 
		\State $Q(\sigma,D,J)\leftarrow \Call{QueryReconstruction}{j,Q(\sigma,D,J)}$
		\State $intermediateResults, Statistics\leftarrow ConstructAndExecute(j)$ \Comment{collect statistical sketches on intermediate data and integrate them on the statistics collection framework}
	    \State $J\leftarrow$ joins in Q(D)
    \EndWhile\\
    
	\State $j\leftarrow \Call{Planner}{J, Statistics}$
	\State \textbf{return}  ConstructAndExecute(j) \\
	
	\Function{PushDownPredicates}{$d,P$}
	   \State $Q(\sigma,\{d\},\emptyset)\leftarrow$ query consists only of $d$ with its local predicates\Comment{$\sigma$ is filled by fields participating in joins}
	   \State $d', Statistics\leftarrow$ Execute(Q($\sigma,\{d\},\emptyset$))\Comment{update original Statistics with the sketches collected for the new $d$}
	   \State \textbf{return} $d'$
	\EndFunction \\
	
	\Function{Planner}{$J,Statistics$}
	  \State $minJoin\leftarrow \emptyset$, $finalJoin\leftarrow \emptyset$
	  \For{$j$ in $J$}
	    \State minJoin$\leftarrow$ min(minJoin, JoinCardinality($j, Statistics$))
	  \EndFor 
     \If{$|J|=2$}
        \State $finalJoin\leftarrow BestAlgorithm(minJoin)\bowtie BestAlgorithm((J-\{minJoin\}))$
     \Else
        \State $finalJoin\leftarrow BestAlgorithm(minJoin)$
     \EndIf
	  \State \textbf{return} $finalJoin$
	\EndFunction \\
	
	\Function{QueryReconstruction}{$j(d_1,d_2),Q(\sigma,D, J)$}
	  \State $d'\leftarrow CreateDataset(j(d_1,d_2))$
	  \State $D\leftarrow (D\bigcup\{d'\})-\{d_1,d_2\}$
	  \State $J \leftarrow J-\{j(d_1,d_2)\}$
	  
	  \State \textbf{return} $Q(\sigma,D, J)$
	\EndFunction 
	
    	\end{algorithmic} 
	\label{algo}
\end{algorithm}

\subsection{Selective Predicates}
\label{predicates}

Filtering can be introduced in the WHERE clause of a query in several forms; here we are focusing on selection predicates.
In the case that a dataset has only one local selection predicate with fixed value, we exploit the equi-height histogram's benefits. Particularly, depending on the number of buckets that we have predefined for the histogram, the range cardinality estimation can reach high accuracy.

However, for multiple selection predicates or complex predicate(s), the prediction can be very misleading. 
In the case of multiple (fixed value) predicates, traditional optimizers assume predicate independence and thus the total selectivity is computed by multiplying the individual ones.
This approach can easily lead to inaccurate estimations \cite{cords}.
In the absence of values for parameters, and given non-uniformly distributed data (which is the norm in real life), an optimizer cannot make any sort of intelligent prediction of selectivity, thus default values are used as described in \cite{selinger} (e.g. 1/10 for equalities and 1/3 for inequalities).
The same approach is taken for predicates with UDFs \cite{hellerstein}.
Most works dealing with complex predicates \cite{chaudhuri,hellerstein} focus on placing such predicates
in the right order and position within the plan, given that the selectivity of the predicate is provided. In our work, we exploit the INGRES \cite{ingres} approach and we push down the execution of predicates (lines 20-23 of Algorithm \ref{algo}) to acquire accurate cardinalities of the influenced datasets. 

As a complex predicate example consider the following query $Q_1$, where we have four datasets, two of which are filtered with UDFs and then joined with the remaining two. (For simplicity in this example we use UDFs but the same procedure is followed for predicates with parameterized values.)



\begin{lstlisting}[
           language=SQL,
           showspaces=false,
           basicstyle=\ttfamily,
           %numbers=left,
           numberstyle=\tiny,
           commentstyle=\color{gray}
        ]
select A.a
from A, B, C, D 
where udf(A) and A.b = B.b 
and udf(C) and B.c = C.c
and B.d = D.d;
\end{lstlisting}

As indicated in line 21 of Algorithm \ref{algo}, we isolate the datasets enhanced with local filters and we create queries for each one of those similarly to the decomposition technique used in INGRES to create single variable queries. In $Q_1$, datasets $A$ and $C$ will be wrapped around the following single variable queries ($Q_2$ and $Q_3$ accordingly): 

\noindent\begin{minipage}{.45\columnwidth}
\begin{lstlisting}[
           language=SQL,
           showspaces=false,
           basicstyle=\ttfamily,
           %numbers=left,
           numberstyle=\tiny,
           commentstyle=\color{gray}
        ]
select A.a, A.b
from A
where udf(A);
\end{lstlisting}
\end{minipage}\hfill
\begin{minipage}{.45\columnwidth}
\begin{lstlisting}[
           language=SQL,
           showspaces=false,
           basicstyle=\ttfamily,
           %numbers=left,
           numberstyle=\tiny,
           commentstyle=\color{gray}
        ]
select C.c
from C
where udf(C);
\end{lstlisting}
\end{minipage}


Note that in both queries the SELECT clause is defined by attributes that participate in the remaining query (i.e in the projection list, in join predicates, or in any other clause of the main query). Once the query construction is completed, we execute them and we save the intermediate results for future processing from the remaining query. At the same time, we also update the statistics (hyperloglog and quantile sketches) attached to the base unfiltered datasets to depict the new cardinalities.  Once this process is finished, we need to update $Q_1$ with the filtered datasets (line 9 in Algorithm \ref{algo}), meaning removing the UDFs and changing the FROM clause. The final query which will be the input to the looping part of our algorithm (lines 11-18) is illustrated below as $Q'_1$.

\begin{lstlisting}[
           language=SQL,
           showspaces=false,
           basicstyle=\ttfamily,
           showstringspaces=false,
           %numbers=left,
           numberstyle=\tiny,
           commentstyle=\color{gray}
        ]
select A'.a
from A', B, C', D 
where A'.b = B.b and B.c = C'.c
and C'.d = D.d;

\end{lstlisting}



\begin{figure*}[h!]
\includegraphics[width=\textwidth, height=8cm]{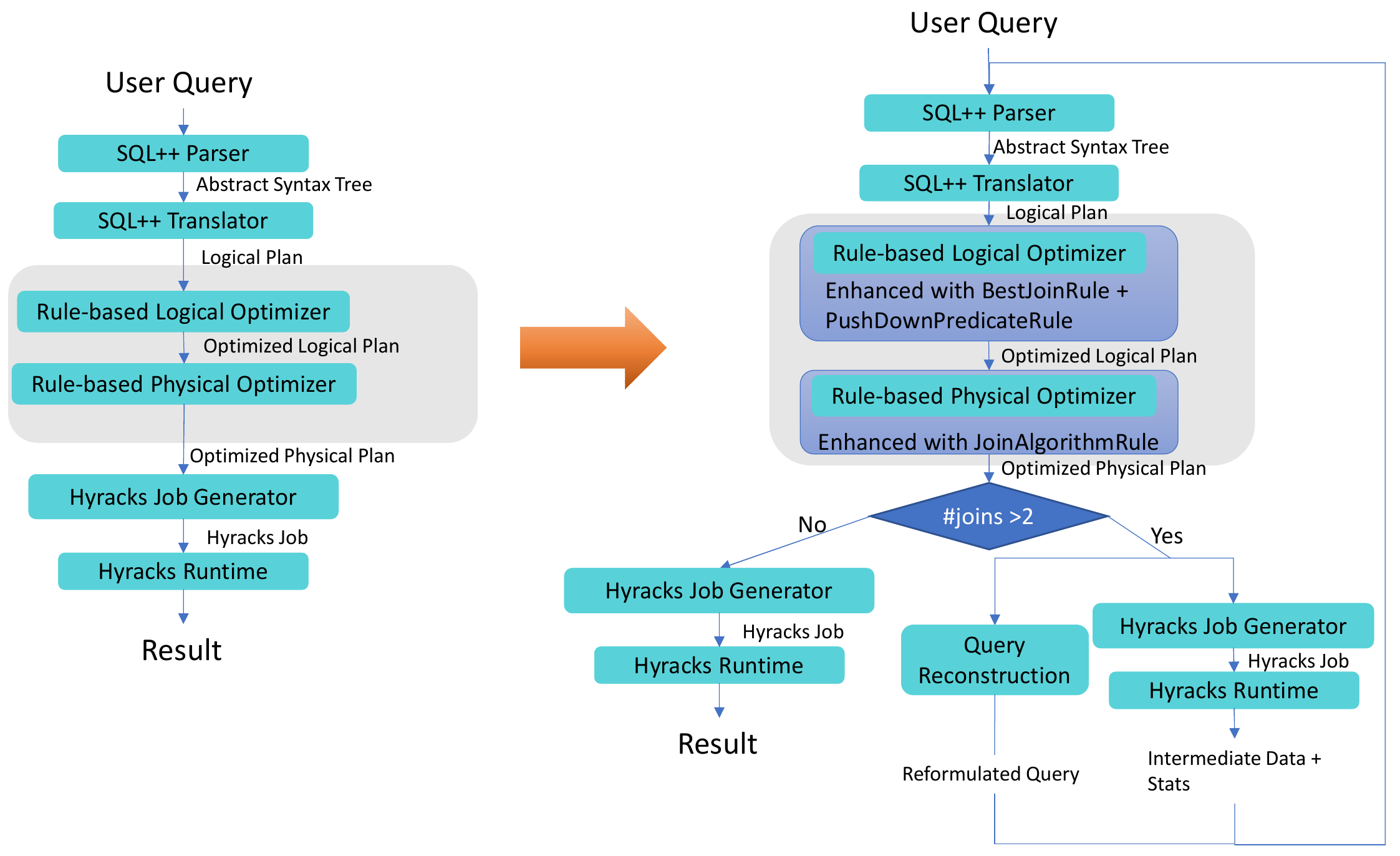}
\caption{AsterixDB workflow without and with the integration of Dynamic Optimization}
\label{asterixdb}
\end{figure*}

\subsection{Planner}
\label{plan}

Next is the Planner stage (lines 25-30), where the input is the non-optimized query (in our case $Q'_1$), along with the most updated statistics. The goal of this stage is to output the best plan (since we focus on joins, this is the plan containing the best join order and join algorithm). 

The first step in the Planner phase is to identify the join with the least result cardinality, along with its algorithm (lines 27-28). After that, we need to construct the join which will be output. If there are more than two joins in the input, then the cheapest join is the output and we are done (lines 31-32). However, in the case that there are only two joins, the Planner will pick the most suitable algorithm for both joins. Then, it will combine the two joins by ordering them according to their result cardinality estimation (lines 29-30 of Algorithm \ref{algo}).

In $Q'_1$ there are three joins, which means that the first case is applied and it suffices to find the cheapest join according to statistics. Assuming that according to formula \ref{join}, A\textquotesingle{} and B lead to the smallest result cardinality, and  A\textquotesingle{} (after the UDF application) is small enough to be broadcast, the plan output is a broadcast algorithm between A\textquotesingle{} and B ($J_{A'B}$).

\subsection{Job Construction}

Next, we construct a job for the plan (in our example, $J_{A'B}$) output by the previous stage (lines 14 and 18 of Algorithm \ref{algo}). 
The details of how we construct a job in AsterixDB are described in section \ref{jc}. The way a job is executed depends on the number of joins in the plan. If there is only one join, it means that we are still inside the looping part of the algorithm (line 14). To that end, we need to materialize the intermediate results of the job and at the same time gather statistics for them. In our example, plan $J_{A'B}$ has only one join - thereby the aforementioned procedure will be followed and the joined results of A\textquotesingle{} and B will be saved for future processing along with their statistics. 

On the other hand, if the plan consists of two joins, it means that the dynamic optimization algorithm has been completed and the results of the job executed are returned back to the user (line 18 of Algorithm \ref{algo}). 

\textbf{Online Statistics: } For the statistics acquired by intermediate results, we use the same type of statistics as described in section \ref{col}. We only gather statistics on attributes that participate on subsequent join stages (and thus avoid collecting unnecessary information).
The online statistics framework is enabled in all the iterations except for the last one (i.e. the number of remaining datasets is three) since we know that we are not going to further re-optimize.

\subsection{Query Reconstruction}
\label{qr}

The final step of the iterative approach is the reconstruction of the remaining query (line 13 of Algorithm \ref{algo}). Given that there will be more re-optimization points (more than two joins remaining), we need to reformulate the remaining query since the part that participates in the job to be executed needs to be removed. 
The following issues need to be considered in this stage:
\begin{itemize}
    \item The datasets participating in the output plan need to be removed (as they are not going to participate in the query anymore) and replaced by the intermediate joined result (lines 36-37).
    \item The join output by Planner needs to be removed (line 38).
    \item Any other clause of the original query influenced by the results of the job just constructed, needs to be reconstructed. 
\end{itemize}

Following our example, the Planner has picked as optimal the join between A\textquotesingle{} and B datasets.  
Consequently this join  is executed first; then, the joined result is stored for further processing and is represented by a new dataset that we call $I_{AB}$.
In terms of the initial query, this will trigger changes in all its clauses. Particularly, in the select clause the projected column derives from one of the datasets participated in the subjob (A). Hence, after its execution, the projected column will now derive from the newly created dataset $I_{AB}$. In the FROM clause both A and B should be removed and replaced by $I_{AB}$. Finally, in the WHERE clause, the join executed has to be removed and if its result participates in any of the subsequent joins, a suitable adjustment has to be made. To this end, in our example B is joined with C in its c attribute. However, the c column is now part of $I_{AB}$. As a result, $I_{AB}$ will now be joined with C. After these changes the reformatted query will look like this ($Q_4$):

\begin{lstlisting}[
           language=SQL,
           showspaces=false,
           basicstyle=\ttfamily,
           %numbers=left,
           numberstyle=\tiny,
           commentstyle=\color{gray},
           mathescape
        ]
select $I_{AB}$.a
from $I_{AB}$, C, D 
where $I_{AB}$.c = C.c and C.d = D.d;

\end{lstlisting}


$Q_4$ has only two joins, which means that the looping part of our algorithm has been completed and that once the Planner picks the optimal join order and algorithm the final job will be constructed and executed with its results returned to the user.

\subsection{Discussion}

By integrating multiple re-optimization points during mid-query execution and allowing complex predicate pre-processing, our dynamic optimization approach can lead to much more accurate statistics and efficient query plans. Nevertheless, stopping the query before each re-optimization point and gathering online statistics to refine the remaining plan introduces some overhead. As we will see in the experimental section, this overhead is not significant and the benefits brought by the dynamic approach (i.e., avoiding a bad plan) exceed it by far. 
Note that here we focus on simple UDF predicates applied on the base datasets. For more expensive UDF predicates, plans that pull up their evaluation need to be considered \cite{hellerstein}. Another interesting point unlocked by dynamic optimization is the forming of bushy join plans. Although they are considered to be expensive as both inputs of the join need to be constructed before the join begins in a parallel environment, they tend to be very efficient as they can open opportunities for smaller intermediate join results.

\section{Integration into AsterixDB}
\label{Integration}

As AsterixDB is supported by two other frameworks (Algebricks and Hyracks), there were multiple changes needed so as to integrate the dynamic optimization approach. 
The left side of Figure \ref{asterixdb} represents the current query processing workflow of the AsterixDB framework, while the right side summarizes our changes. In particular, in the beginning the workflow behaves in the same way as always, with the exception of few additional rules integrated into the rule-based (JoinReOrderRule, PushDownPredicateRule) and physical-based (JoinAlgorithmRule) optimizer (\textbf{Planner}). Afterwards, depending on the number of joins participating in the query currently being processed, we either construct and execute the Hyracks job and output the result to the user as usual (only two joins) or we perform the following two steps (more than two joins): 
\begin{itemize}
    \item We introduce the \textbf{Query Reconstruction} phase where we reformulate the query currently being processed and we redirect it as new input to the SQL++ parser and the whole query process starts from the beginning once again.
    \item We construct a Hyracks job (\textbf{Job Construction}) by using various new operators introduced to allow materialization of the results of the query currently being processed along with connection of previously (if any) executed jobs. 
\end{itemize}

\subsection{Planner}

 If a dataset has more than one filter, the PushDownPredicateRule is triggered. This rule will push the filters down to their datasource and will remove the rest of the operators from the plan, leading to a modified plan of a simple select-project query (like $Q_2$ and $Q_3$ in section \ref{predicates}) . 
On the other hand, if there is only one filter, we estimate the filtered dataset cardinality based on histograms built on the base dataset.

Afterwards, the Planner stage will decide the optimal join order and algorithm. In order for the Planner to pick the join with the least cardinality, we enhanced the rule-based logical Optimizer (part of the Algebricks framework) with the JoinReOrderRule (see Figure \ref{asterixdb}). 
To further improve the efficiency of the execution plan, we integrated a rule in the rule-based physical Optimizer (Figure \ref{asterixdb}) that picks the most suitable join algorithm.

\subsubsection{Join Ordering}
\leavevmode

\begin{figure}[h!]
\includegraphics[width=0.7\columnwidth]{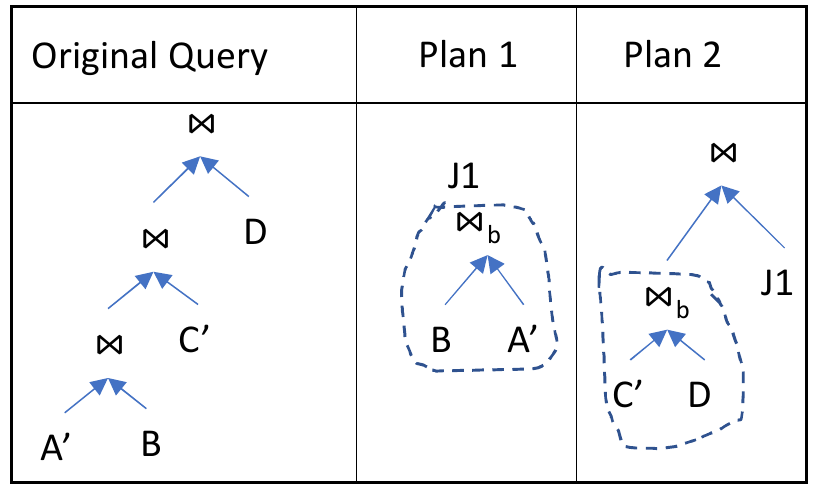}
\caption{Planning Phase when Dynamic Optimization is triggered}
\label{plans}
\end{figure}

The main goal of the join order rule is to figure out the join with the least cardinality. To that end, we identify all the individual joins along with the datasources (post-predicate execution) of their predicates. In this work, we focus only on joins as formed in the WHERE clause of the query. In the future, we plan to infer more possible joins according to correlations between join predicates. Afterwards, we apply formula \ref{join} based on statistics (see Section \ref{col}) collected for the datasets and predicates involved in the join.
Traditional optimizers that are based on static cost-based optimization need to form the complete plan from the beginning, meaning that we need to search among all different possible combinations of joins which can be very expensive depending on the number of base datasets. However, in the case of incremental optimization, it suffices to search for the cheapest join because the rest will be taken into consideration in the next iterations of our algorithm. In our example in Figure \ref{plans}, in $Q_1$ the join between post-predicate $A$ ($A$\textquotesingle{}) and $B$ will be estimated as the cheapest one and will be output from the Planner stage.  

The second feature of this rule is triggered when there are only two joins left in the query and hence the statistics obtained up to that point suffice to figure out the best join order between them. Specifically as depicted in Plan 2 of Figure \ref{plans}, in this case a two-way join (between three datasets) is constructed whose inputs are (1) the join (between two of the three datasets) with the least result size (estimated as described above) and (2) the remaining dataset. 

It is worth noticing that in the first iteration of the approach the datasets that are joined are always among the base datasets. However, in the rest of the iterations, one or both of the joined datasets may be among the results from previous iterations. An example of that is shown in Plan 2 of Figure \ref{plans}, where the right dataset of the final join is the result of the first iteration (J1) of our algorithm. 

\subsubsection{Join Algorithm}
\leavevmode

While hash join is the default algorithm, by having accurate information about the datasets participating in the corresponding join, the optimizer can make more efficient decisions. If one of the datasets is small enough, like A\textquotesingle{} and C\textquotesingle{} in our example (see Figure \ref{plans}), then it can be faster to broadcast the whole dataset and avoid potential reshuffling of a large dataset over the network.

Knowing that the cardinality of one of the datasets is small enough to be broadcast also opens opportunities for performing the indexed nested loop join algorithm as well. However, two more conditions are necessary to trigger this join algorithm. The first one is the presence of a secondary index on the join predicate of the "probe" side. The second condition refers to the case of primary/foreign key join and dictates that the dataset that gets broadcast must be filtered - thereby during the index lookup of a large dataset there will be no need for all the pages to be accessed. 

\subsection{Query Reconstruction}

This stage is entered in one of the following cases: (1) the Planner has output a simple projection plan (predicate push down) or (2) the Planner output is a select-project-join plan (cheapest join). In both cases, we follow the process described in section \ref{qr} to reformulate the clauses of the input query and output the new query that will be given as input to the optimizer for the remaining iterations of our algorithm. 

\subsection{Job Construction}
\label{jc}
There are three different settings when creating a job: 
\begin{enumerate}
    \item When there are still re-optimizations to be scheduled (more than 2 joins), the output of the job has to be materialized for future use.
    \item If one or both inputs of a job is a previously materialized job output, we need to form a connection between the jobs.
    \item When the iterations are completed, the result of the last job
    will be returned to the user. 
\end{enumerate}
 
\begin{figure}[h!]
\includegraphics[width=\columnwidth, height=6.5cm]{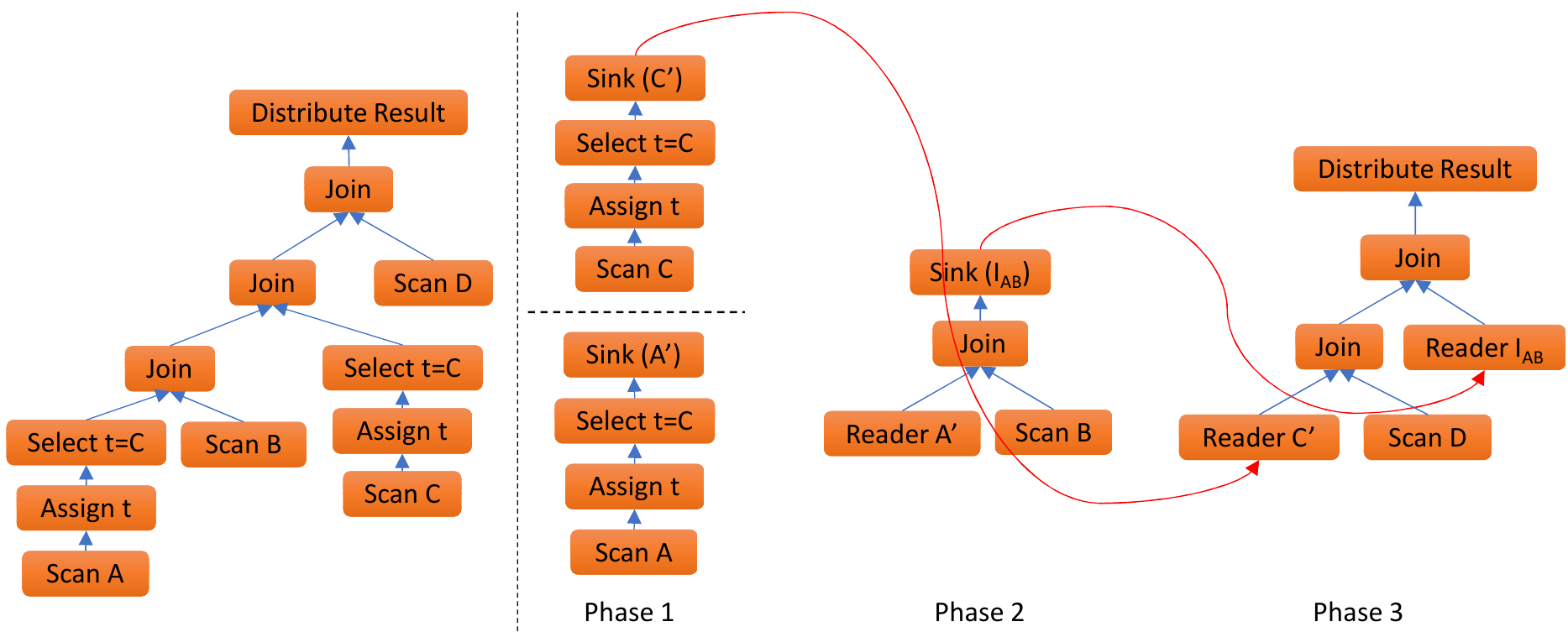}
\caption{Original Hyracks job split into smaller jobs}
\label{splitting}
\end{figure}

We use the example in Figure \ref{splitting} to illustrate the process we followed to satisfy the aforementioned cases.
The left side of the figure 
depicts the usual job for the three-way join query ($Q_1$), where the final result is returned to the user via the DistributeResult operator. 
Instead, on the right side of the Figure (Phase 1), two subjobs are created which push down the UDF predicates applied to datasources A and C. Their results are the post-predicate versions of A and C (Sink(A') and Sink(C') accordingly). The Sink operator is responsible for materializing intermediate data while also gathering statistics on them. 

In Phase 2, the subjob formed wraps the join between datasets A' and B, as this is the plan output by the Planner. Note that the new operator introduced in this phase (Reader A') indicates that a datasource is not a base dataset. Instead, it is intermediate data created by a previous subjob. In our example, Reader A' represents the materialized data created in the previous phase by Sink(A'). Since the original query has not finished yet (remaining joins), the Sink operator will be triggered once again and it will store in a temporary file the joined results ($I_{AB}$), while at the same time it will collect the corresponding statistics.

Finally, the goal of Phase 3 is to wrap the output of the Planner which is a two-way join. The existence of two joins indicates that we are at the final iteration of the dynamic approach - thereby this job is the final one and its result should be returned to the user. Consequently, the DistributeResult operator re-appears in the job, as depicted in Figure \ref{splitting}. 

\subsection{Discussion}

To integrate the dynamic optimization approach in the AsterixDB framework, we had to create an iterative workflow which gave us the opportunity to trigger multiple re-optimization points that result in more efficient query plans. 
In this work, we concentrate on multi-join queries which may also contain multiple and/or complex selection predicates. Although other types of operators may exist in the query, for now they are evaluated after all the joins and selections have been completed and traditional optimization has been applied. 
In the future, we plan to investigate more costly UDF predicates that may instead be better to be pulled up for evaluation.

\begin{figure*}[h!]
\includegraphics[width=\textwidth]{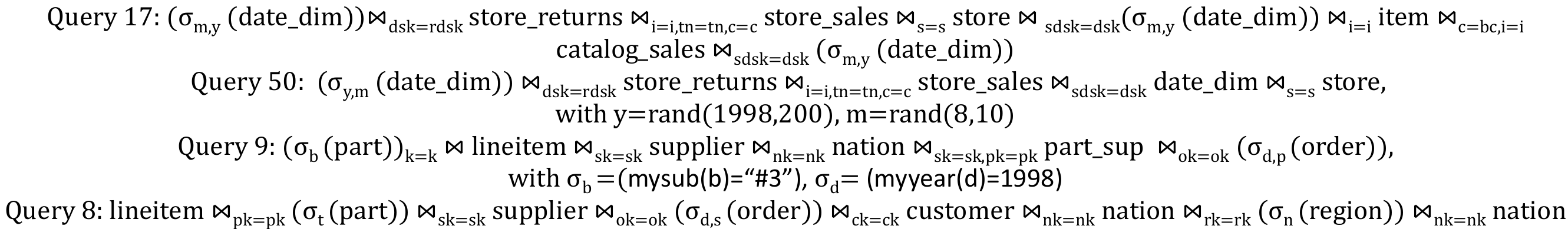}
\caption{Queries used for the experimental comparisons.}
\label{queries}
\end{figure*}

\begin{figure*}[h!]
\includegraphics[width=\textwidth]{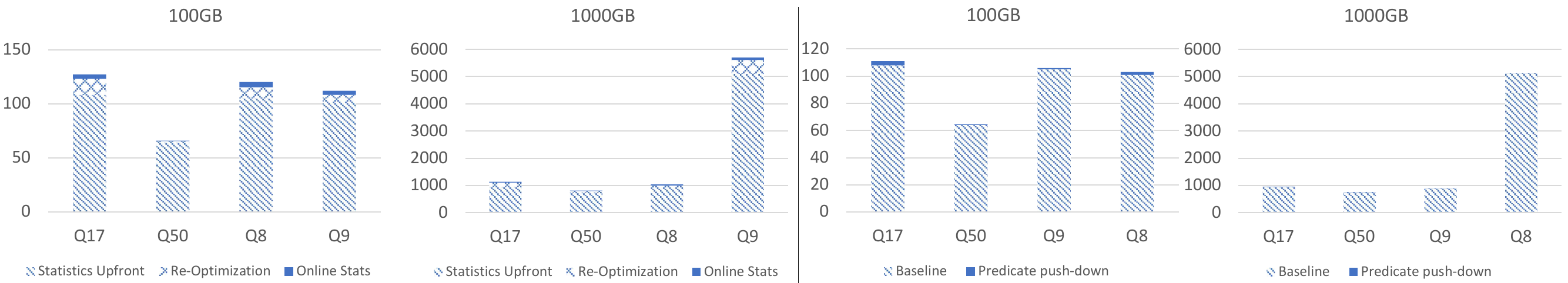}
\caption{Overhead imposed by the multiple re-optimization points and the online statistics.}
\label{overhead}
\end{figure*}


\section{Experimental Evaluation}
\label{Experiments}

We proceed with the performance evaluation of our proposed strategies and discuss the related trade-offs. The goals of our experiments are to: (1) evaluate the overheads associated with the materialize and aggregate statistics steps; (2) show that good join orders and methods can be accurately determined, and (3) exhibit the superior performance and accuracy over traditional optimizations. In particular, in the following experiments we compare the performance of our dynamic approach with: (i) AsterixDB with the worst-order, (ii) AsterixDB with the best-order (as submitted by the user), (iii) AsterixDB with static cost-based optimization, (iv) the pilot-run \cite{pilot} approach, and (v) an INGRES-like approach \cite{ingres}. Section \ref{execution} contains detailed explanations of each optimization approach. 

\textbf{Experimental Configuration:} All experiments were carried out on a cluster of 10 AWS nodes, each with an Intel(R) Xeon(R) E5-2686 v4 @ 2.30GHz CPU (4cores), 16GB of RAM and 2TB SSD. The operating system is 64-bit Red-Hat 8.2.0. Every experiment was carried out five times and we calculated the average of the results.

\textbf{Queries:} We evaluate the performance using four representative queries from TPC-DS (Query 17 and Query 50) \cite{tpcds} and TPC-H \cite{tpch} (Query 8 and Query 9). The actual queries are shown in Figure \ref{queries}.
These queries were selected because of: (1) their complexity (from the number of joins perspective), and, (2) their variety in join conditions (primary/foreign key vs fact-to-fact joins). 

To better assess the effect of selection predicates on our runtime dynamic approach, we used modified versions of Queries 8, 9 and 50. Specifically, to consider multiple fixed value predicates, in Query 8 we added two (and correlated \cite{automatic}) predicates on the \textit{orders} table.
We use Query 9 to examine the effect of UDFs (by adding  various UDFs on top of the \textit{part} and \textit{orders} tables.
Finally, in Query 50,  we added two selections with parameterized values on top of one of the dimension tables. (The SQL++ version of all the queries appears in the Appendix.) 

For all of the scenarios we generate 3 TPC-DS and 3 TPC-H datasets with scale factors 10, 100, 1000. A scale factor of 1000 means that the cumulative size for the datasets involved in the specific query is 1TB. All the data is directly generated and then loaded into AsterixDB. 
It is also worth noting that we gain upfront statistics for the forming of the initial plan during the loading of the datasets in AsterixDB. This is only performed once and it is not part of the query execution process; thus the performance numbers reported in our results do not include that part. The loading times can vary from 10 minutes to 8 hours depending on the size of the datasets. However, as was shown in \cite{ildar}, the statistics collection overhead is minimal with respect to the loading time.


\subsection{Overhead Considerations}

In this section, we evaluate the overhead introduced to the AsterixDB execution time due to our dynamic optimization techniques, namely (1) the introduction of re-optimization points, (2) the gathering of statistics during runtime, and (3) the separate execution of multiple/complex predicates. To this end, we report the execution times for the above four representative queries for scale factors 100 and 1000. 

For the first two settings we perform the following three executions for each query. In the first execution we acquired all the statistics needed for forming the optimal execution plan by running our runtime dynamic optimization technique. Then, we re-executed the query by having the updated statistics for each dataset so that the optimal plan is found from the beginning. In the final execution, we enabled the re-optimization points but we removed the online statistics collection. That helped us assess the overhead coming from writing and reading materialized data. Finally, to evaluate the cost of online statistics gathering we simply deducted the third execution time (re-optimization) from the first one (whole dynamic optimization technique). 

As seen in the left side of figure \ref{overhead}, for scale factor 100, the total re-optimization time is around 10\% of the execution time for most queries, with the exception of Q50 which has only four joins leading to an overhead of 2\%. Particularly, the four joins introduce two re-optimization points before the remaining query has only two joins and there is no need for further re-optimization. There is also a re-optimization in the beginning of this query introduced by the execution of the filtered dataset. However, this is insignificant as will be discussed later.
For the scale factor of 1000, the  overhead of re-optimization increases up to 15\% for most queries, as the intermediate data produced are larger and thus the I/O cost introduced by reading and writing intermediate data is increased.

The online statistics collection brings a small overhead of 1\% to 3\% (scale factor 100) to the total execution time, as it is masked from the time we need to store and scan the intermediate data. Moreover, the extra time for statistics depends on the number of attributes for which we need to keep statistics for. Following the example of Q50 as above, the statistics collection overhead is only 1\% because it has the smallest number of join conditions. In scale factor 1000 the overhead of gathering statistics is increased, as the data upon which we collect statistics are larger in size, but it remains insignificant (up to 5\%). Overall, we observe a total of 7-13\% overhead for scale factor 100 and up to 20\% for scale factor 1000. We believe that this is acceptable given the beneﬁts brought by our approach, as will be shown in Section \ref{execution}.

Finally, we assess the overhead of applying the incremental optimization approach to estimate the influences of multiple/complex predicates. For the base setup, we deactivated the multiple re-optimization points and executed the plan formed as if the right statistical data is available from the beginning. 
Then, the experiment was repeated by enabling the dynamic optimization only for materializing the intermediate results coming from pushing down and executing multiple predicates. 
The remaining query was executed based on the refined statistics coming from the latter step. As the results show (right side of figure \ref{overhead}), even in the case of Q17, where there are multiple filters present, the overhead does not exceed  3\% of the total execution time, even for scale factor 1000. On the other hand, Q50 once again has the smallest overhead as there is only one dataset filtered.

\begin{figure*}[t!]
\includegraphics[width=\textwidth]{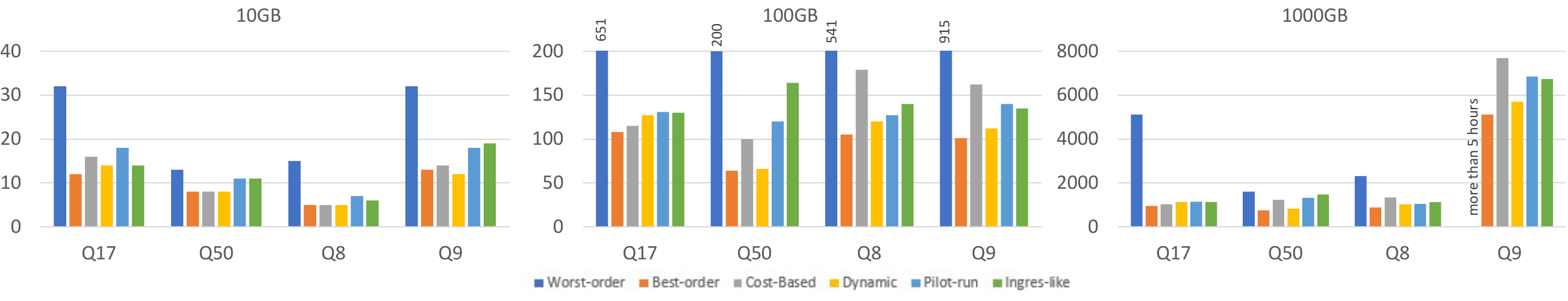}
\caption{Comparison between Dynamic Optimization, traditional cost-based optimization, regular AsterixDB ( join best-order vs worst-order), Pilot-run and Ingres-like }
\label{exp}
\end{figure*}

\subsection{Comparison of Execution Times}
\label{execution}

We proceed to evaluate our dynamic approach techniques against: (i) the join worst-order, (ii) the join best-order, (iii) a traditional cost-based optimization and (iv) the pilot-run method \cite{pilot}. 
For the worst-order plan, we enforce a right-deep tree plan that schedules the joins in decreasing order of join result sizes(the size of the join results was computed during our optimization). 
The best-order plan assumes that the user knows the optimal order generated by our approach and uses that order in the FROM clause when writing the query. We also put some broadcast hints so the default optimizer can choose the broadcast algorithm.
These two settings represent the least and the most gain, accordingly, that we can achieve with our approach against the default approaches of AsterixDB. 

To compare with a traditional cost-based optimization approach, we collected statistics on the base datasets during the ingestion phase and we formed the complete execution plan at the beginning based on the collected statistics. When UDFs or parameters are present in a query we use the default selectivity factors as described in \cite{selinger}.
For the pilot-run method, we gathered the initial statistics by running select-project queries (pilot-runs) on a sample of each of the base datasets participating in the submitted query. If there are predicates local to the datasets, they are included in the pilot-runs. In the sampling technique used in \cite{pilot} during pilot runs, after $k$ tuples have been output the job stops. To simulate that technique we enhanced our "pilot runs" with a LIMIT clause. Based on those statistics, an initial (complete) plan is formed and the execution of the original query begins until the next re-optimization point where the plan will be adjusted according to feedback acquired by online statistics. 

Finally, for the INGRES-like approach we use the same approach as ours to decompose the initial query to single variable queries. However, the choice of the next best subquery to be executed is only based on dataset cardinalities (without other statistical information). Furthermore, in the original INGRES approach intermediate data are stored into a new relation;  in our case we store it in a temporary file for simplicity. 
The experimental results 
are shown in Figure \ref{exp}. (The actual query plans produced for each query for this and later experiments appear in the Appendix).

\subsubsection{TPC-DS}
\leavevmode

\textit{Query 17:} This query has a total of 8 base tables (Figure \ref{queries}). Three of those (i.e. dimension tables) are attached to selective filters and are used to prune down the three large fact tables, while \textit{item} and \textit{store} (i.e. smaller tables) are used for the construction of the final result. Our dynamic optimization approach will find that the optimal plan is a bushy tree, as dimension tables should be joined with the fact tables to prune down as much as possible the intermediate data. Then, they will be joined with each other to form the result. It is also worth noting that our approach will find that the dimension tables and \textit{store} will be broadcast in all scale factors along with \textit{item} in factors 10 and 100. 

Given that there are no complex predicates, all other approaches (apart from the worst-order) will form similar bushy trees along with the suitable join algorithm in the appropriate cases. Hence, our dynamic optimization approach does not bring any further benefit (in fact there is a slight degradation, around 1.15-1.20x depending on the scale factor, against best-order due to the overhead introduced by re-optimization). Finally, the worst-order will join the fact tables first, resulting in very large intermediate results and a 5x slower performance.

\textit{Query 50:} This query contains two dimension tables (\textit{date\_dim}) only one of which is filtered (with parameterized expressions), two large tables and \textit{Store} that helps pruning down the final result. The optimal plan found by our dynamic approach first prunes down one of the fact tables by joining it with the filtered dimension table and then joins it with the other large table. Our approach is also able to choose the broadcast algorithm whenever appropriate. With the enhancement of broadcast hints, best-order will pick exactly the same execution plan, leading to slightly better performance than our dynamic approach (1.05, 1.1x for scale factors 100 and 1000). 

Cost-based optimization results in a different plan because of the inaccurate cardinality estimates on the post-filtered dimension table and on the joined result between the fact tables. As a result, although it finds most of the broadcast joins, it leads to a 1.5x worse performance than our approach for scale factors 100 and 1000. A bushy tree will be formed by the INGRES-like approach due to its naive cost-model approach (considering only dataset cardinalities), resulting in an even worse performance. The worst-order of AsterixDB will trigger hash joins by default. On top of that, it will schedule the join between the fact tables in the beginning; thus it has the worst performance. Lastly, pilot-run makes the wrong decision concerning the join ordering between the large tables because of inaccurate statistics and thereby is around 1.8x slower than our approach.

\subsubsection{TPC-H}
\leavevmode

\textit{Query 9:} The lineitem table is joined on foreign/primary key with four smaller tables and on foreign key with \textit{part\_sup}. Once again, our approach will find the optimal plan, which in this case is a bushy tree. Apart from the correct join-order, our techniques will pick the broadcast algorithm in the case of the \textit{part} table for scale factors 10 and 100, as well as in the case of the joined result of \textit{nation} and \textit{supplier} tables. Cost-based optimization will find a similar bushy tree; however, due to wrong cardinality estimation, it will not broadcast the part table and the intermediate data produced by joining nation and supplier will only be broadcast for scale factor 10. As a result, our approach has a slightly better performance than the cost-based one. Similarly, the best-order will form the optimal execution plan leading to the best performance once again. 

As with all the other queries, the worst-order will schedule the largest result producing joins in the beginning along with the hash algorithm, which will result in an execution time more than 5 hours. Hence, almost all techniques were 7x better than the worst-order. In the pilot-run case, once again, a suboptimal plan is chosen due to inaccurate unique cardinalities estimated by initial sampling. Finally, once again the INGRES-like approach will form a less efficient bushy tree since it focuses only on dataset cardinalities. 

\begin{figure*}[t!]
\includegraphics[width=\textwidth]{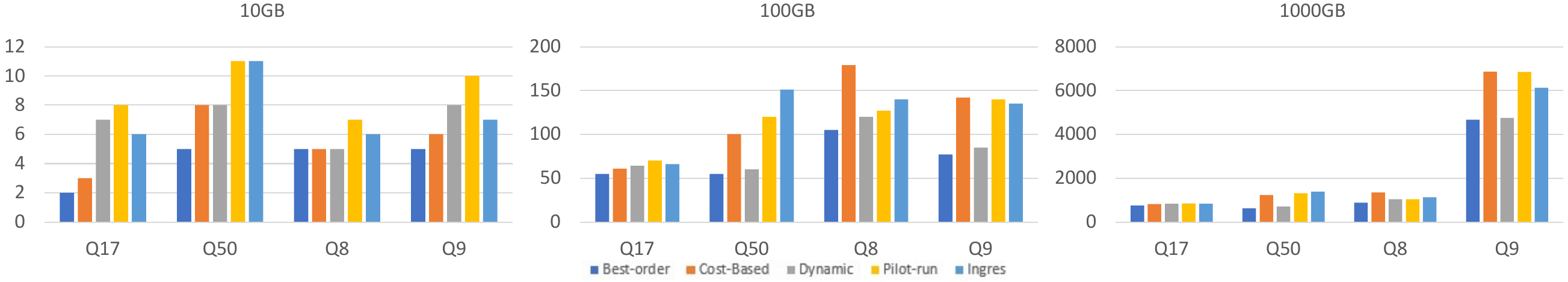}
\caption{Comparison between Dynamic Optimization, traditional cost-based optimization, regular AsterixDB ( join best-order vs worst-order), pilot-run and ingres-like when INL join is considered.}
\label{udfs}
\end{figure*}

\textit{Query 8:} This query has eight datasets in total. The lineitem table is a large fact table while all the others are smaller (three of them are filtered with multiple predicates). All the joins between the tables are between foreign/primary keys. 
Again our approach manages to find the optimal plan (bushy join) as it uses the dynamic optimization techniques described above to calculate the sizes of base datasets after multiple-predicate filters are applied. The dynamic approach also gives the opportunity to the optimizer to choose the broadcast algorithm when appropriate, mainly for scale factors 10 and 100. Best-order will form the same execution plan (both in terms of join order and algorithm) as the dynamic approach and it will be more efficient since there is no re-optimization. 

In the cost-based case, due to inaccurately estimated cardinalities on the post-filtered \textit{orders} table, a different bushy plan is chosen. Although for scale factor 1000, the benefit of broadcast opportunities picked by the dynamic approach is not as noticeable as in the rest of the scale factors, it is still 1.3x faster than the cost-based one since it forms a better plan.
Furthermore, pilot-run forms the same optimal plan as our approach, but because of the overhead introduced by pilot runs is slightly slower. The INGRES-like approach will focus only on dataset cardinalities and not on statistical information and thus it will find a suboptimal plan. Finally, the worst-order leads to a right-deep join with hash joins that can be up to 2.5x worse than our approach.



The last set of experiments examine the behavior of our approach when the Indexed Nested loop Join (INLJ) is added as another possible join algorithm choice. 
We thus enhanced the TPC-H and TPC-DS datasets with a few secondary indexes on the attributes that participate in queries as join predicates and are not the primary keys of a dataset. The worst-order is excluded from these experiments since in the absence of hints, it will not choose INL; hence its execution time will not change. 
The results of these experiments are shown in Figure \ref{udfs}. 

\subsubsection{TPC-DS}
\leavevmode

\textit{Query 17:} In this particular query, there are 3 cases where the INL join will be picked by the dynamic approach for all scale factors. All of these cases are for the foreign/primary key joins between the large fact tables and the post-filtered dimension tables. In these particular cases the dimension tables are small enough to be broadcast but at the same time they have been filtered; hence not all pages of the large fact tables satisfy the join and need to be accessed. The same will happen with all the other approaches - thereby the execution time will be better in all cases. To that end, our dynamic approach will not bring any further benefit in this particular case.


\textit{Query 50:} In this query, the dynamic approach will pick the INL join algorithm only in the case of the join between the filtered dimension table and the \textit{store\_returns} table. However, \textit{store\_returns} is not a very large table, and thus scanning it instead of performing an index lookup does not make a big difference; this results in a smaller improvement compared to the performance in the previous section. The INGRES-like approach similar to the dynamic one, will pick the INL join for \textit{store\_returns}$\bowtie$\textit{date\_dim} because \textit{date\_dim} is small enough to be broadcast (after it has been filtered) and \textit{store\_returns} has a secondary index on its join predicate. Finally, pilot-run and cost-based will miss the opportunity for choosing INL since the \textit{store\_returns}  joined with the dimension table and derives from intermediate data; thus the needed secondary index does not exist anymore. Consequently, the difference in the performance against the dynamic approach is even bigger.

\subsubsection{TPC-H}
\leavevmode

\textit{Query 9:} Dynamic optimization leads to the choice of INL for the join between \textit{lineitem} and \textit{part}. Thus, the query executes much faster than in the previous section. The same happens with all other approaches apart from the pilot-run in which, similarly to the previous query,  \textit{lineitem} does not have a secondary index anymore, thus leading to a performance degradation compared to the dynamic approach.

\textit{Query 8:} This is a case where the INL cannot be triggered for any of the approaches. For example, in the cost-based approach, when \textit{lineitem} and \textit{part} are joined, although there is a secondary index on the \textit{lineitem} predicate and \textit{part} is filtered, the latter is not small enough to be broadcast. In the other approaches, in \textit{supplier} $\bowtie$ \textit{nation} the nation does not have a filter on it; hence, although all the other requirements are met, a simple broadcast will be better because scanning the whole dataset once is preferred to performing too many index lookups.  

\subsection{Discussion}

\begin{table}
\begin{center}
\small
 \begin{tabularx}{\columnwidth}{| >{\centering\arraybackslash}X 
  | >{\raggedright\arraybackslash}X 
  | >{\centering\arraybackslash}X 
  | >{\centering\arraybackslash}X 
  | >{\centering\arraybackslash}X 
  | >{\centering\arraybackslash}X |} 
 \hline
 Data Size (GB)	 	&  Cost-Based  & Pilot-run & Ingres-like & Best-order & Worst-order \\ 
 \hline
 \hline
 100 			& 1.34x  & 1.28x  & 1.4x & 0.88x & 5.2x  			\\ 
  \hline
1000 			& 1.27x  & 1.20x  & 1.27x  & 0.85x   & >10x                 \\
 \hline
 \end{tabularx}
 \vspace{1em}
\caption{Average improvement of the runtime dynamic approach against the other optimization methods.}
\label{avg}
\end{center}
\vspace{-3em}
\end{table}

The results of our evaluation showcase the superiority of our dynamic optimization approach against traditional optimization and state-of-the-art techniques. 
Table \ref{avg} shows the average query time improvement of the dynamic approach (among all 5 executions of each of the four queries for each data size).
It is worth mentioning that the best improvement is observed for the 100GB dataset size.
When the base dataset is large enough, a wrong execution plan chosen by traditional optimizers is noticeable and at the same time the broadcast join has a higher possibility of being picked by our approach due to accurate selectivity estimations (post execution of predicates). For the 1000GB dataset size, we observed less improvement with our approach (see Table \ref{avg}), as broadcast joins are limited, and the intermediate results are larger leading to a larger I/O cost. Nevertheless, we were still better than all the other approaches. For the 10GB size, we have the least improvement (there are even cases where we are worse than cost-based) because the base datasets are very small in size and the overhead imposed by the intermediate data materialization is noticeable. A further interesting observation is that most of the optimal plans are bushy joins, meaning that even if both inputs have to be constructed before the join is performed, forming the smaller intermediate join results brings more benefits to the query execution. 

With respect to the overhead derived by our dynamic optimization techniques, we note that although in the worst case (scale factor 1000) the cost can be expensive, in most cases our plans are still faster than the plans produced by traditional optimizers. 
\section{Conclusions}
\label{Conclusion}

In this paper we have investigated the benefits of using dynamic query optimization in big data management systems. We described how we decompose a submitted query into several subqueries with the ultimate goal of integrating re-optimization points to gather statistics on intermediate data and refine the plan for the remaining query. Although our work concentrates on complex join queries, we also treat multiple selective predicates and predicates with parameterized values and UDFs, as part of the re-optimization process. That way, in addition to the benefit of gathering information about the cardinality of intermediate data, we also get more accurate estimations about the sizes of filtered base datasets. We chose AsterixDB to implement our techniques as it is a scalable BDMS optimized to execute joins in a pipeline. We were able to showcase that, even though it blocks the pipelining feature and introduces intermediate results, our approach still gives amost always the best performance. 

We evaluated our work by measuring the execution time of different queries and comparing our techniques against traditional static cost-based optimization and the default AsterixDB query execution approach and we proved its superiority. 
When querying big data, it pays to get good statistics by allowing re-optimization points since a small error in estimating the size of a big dataset can have much more drastic consequences on query performance than the overhead introduced. Nevertheless, our approach performs at its best when complex predicates are applied to the base datasets of a query or the join conditions are between fact tables (leading to skewness in selectivity and join result estimation accordingly). 

In future research we wish to explore ways to address more complex UDFs in our dynamic optimization approach. Further, we want to exploit the benefits of dynamic optimization when other operators (i.e group-by, order by, etc.) are included in the query. Although more re-optimization points make our technique more accurate and robust, they also increase its overhead. Consequently, it would be interesting to explore (through a cost model) the trade-off of facilitating the dynamic optimization approach but with fewer re-optimizations and still obtain accurate results. Finally, runtime dynamic optimization can also be used as a way to achieve fault-tolerance by integrating checkpoints. That would help the system to recover from a failure by not having to start over from the beginning of a long-running query. 

\bibliographystyle{ACM-Reference-Format}
\bibliography{references}


\section{Appendix}
In this section, we provide the SQL++ version of the queries discussed in the experimental section. 
Moreover, we provide the detailed plans generated by the different optimizers for the queries in Section \ref{execution}. The $\bowtie$ join represents a hash-based join unless it is marked with \textbf{`b'} which denotes a broadcast join or \textbf{'i'} which denotes a indexed nested loop join.

\begin{figure}[h]
  \begin{subfigure}[b]{\columnwidth}
    \includegraphics[width=\linewidth]{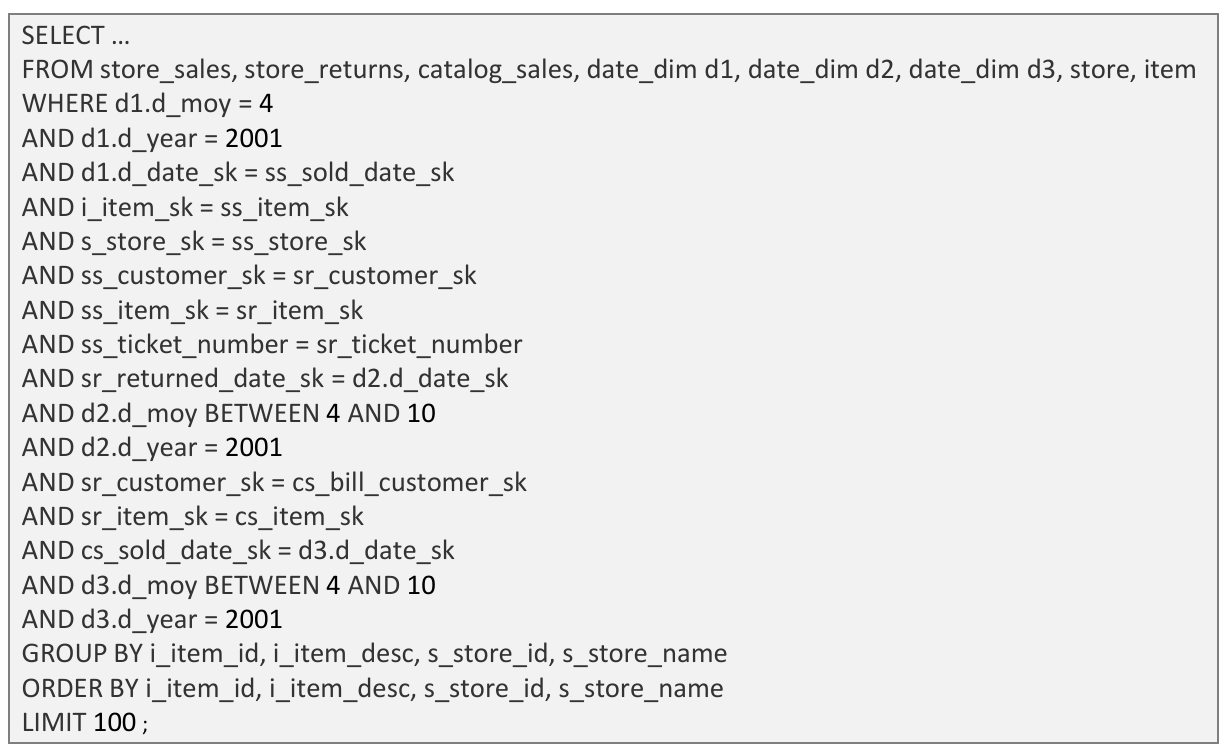}
    \caption{Query 17}
  \end{subfigure}
  \begin{subfigure}[b]{\columnwidth}
    \includegraphics[width=\linewidth]{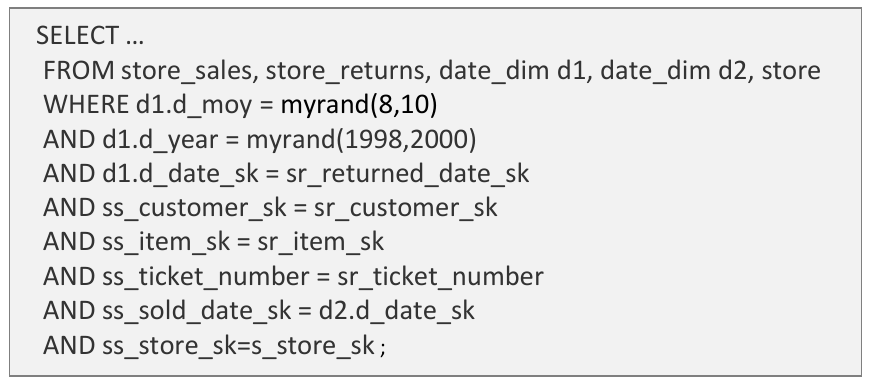}
    \caption{Query 50}
  \end{subfigure}
  \caption{TPC-DS Queries: (a) 17 and (b) 50.}
  \label{plans_udfs}
\end{figure}

\begin{figure}[h]
  \begin{subfigure}[b]{\columnwidth}
    \includegraphics[width=\linewidth]{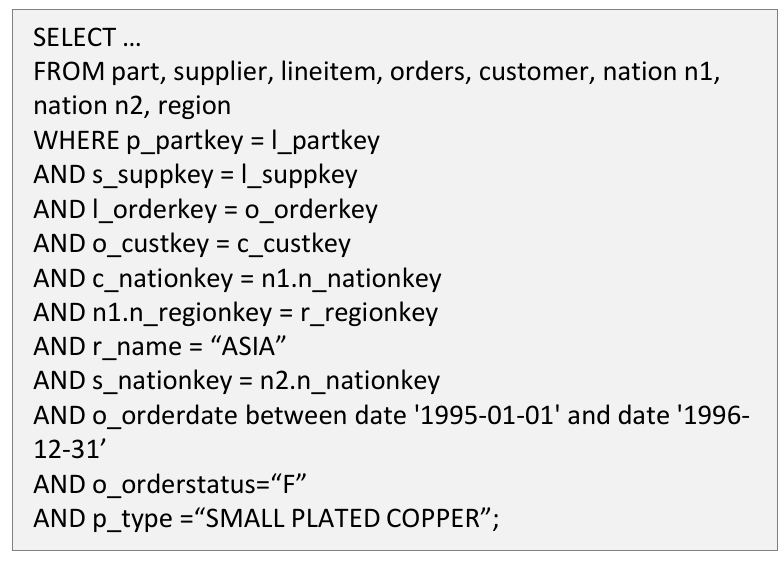}
    \caption{Query 8}
  \end{subfigure}
  \begin{subfigure}[b]{\columnwidth}
    \includegraphics[width=\linewidth]{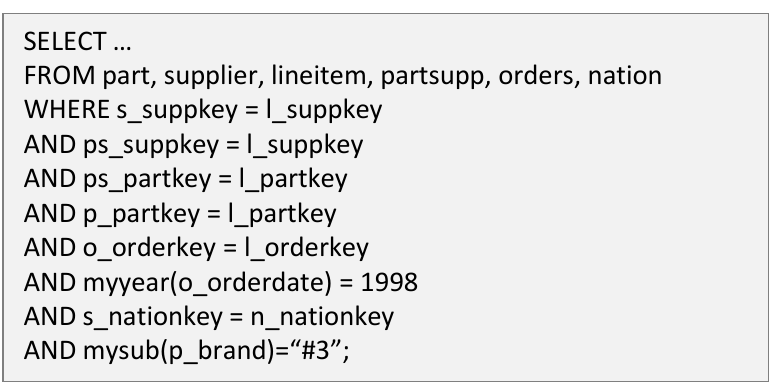}
    \caption{Query 9}
  \end{subfigure}
  \caption{TPC-H Queries: (a) 8 and (b) 9.}
  \label{plans_udfs}
\end{figure}

\begin{figure*}[h!]
\includegraphics[width=\textwidth]{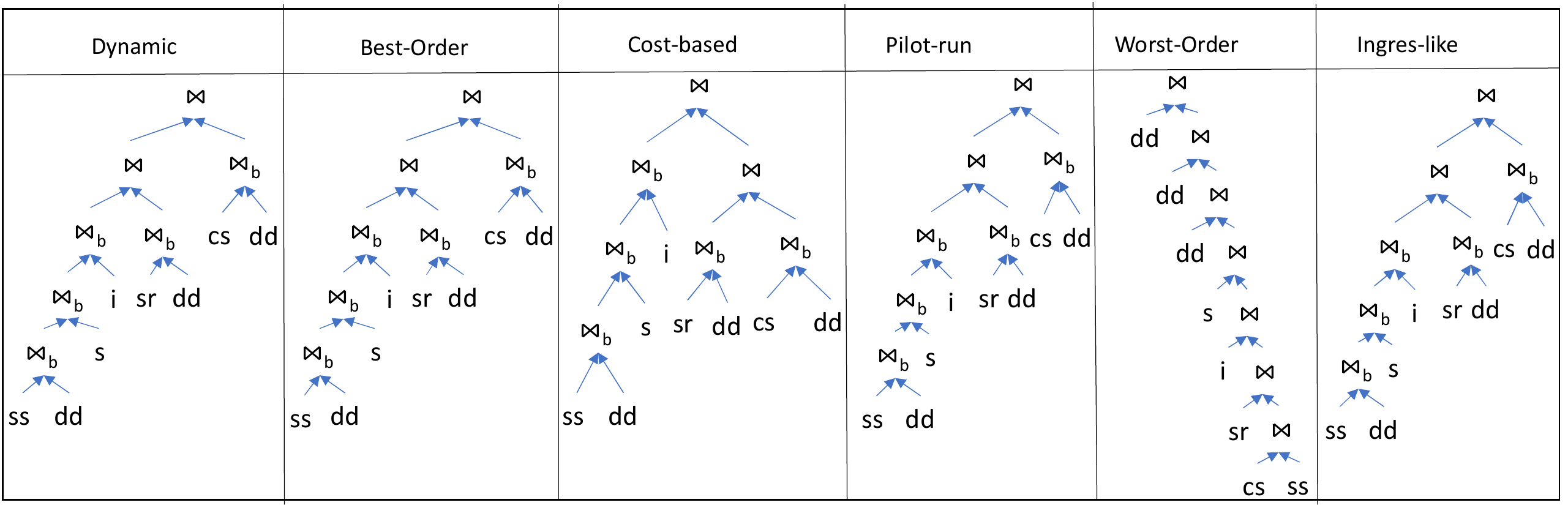}
\caption{Plans Generated for Query 17, Figure \ref{exp}, 10GB.}
\end{figure*}

\begin{figure*}[h!]
\includegraphics[width=\textwidth]{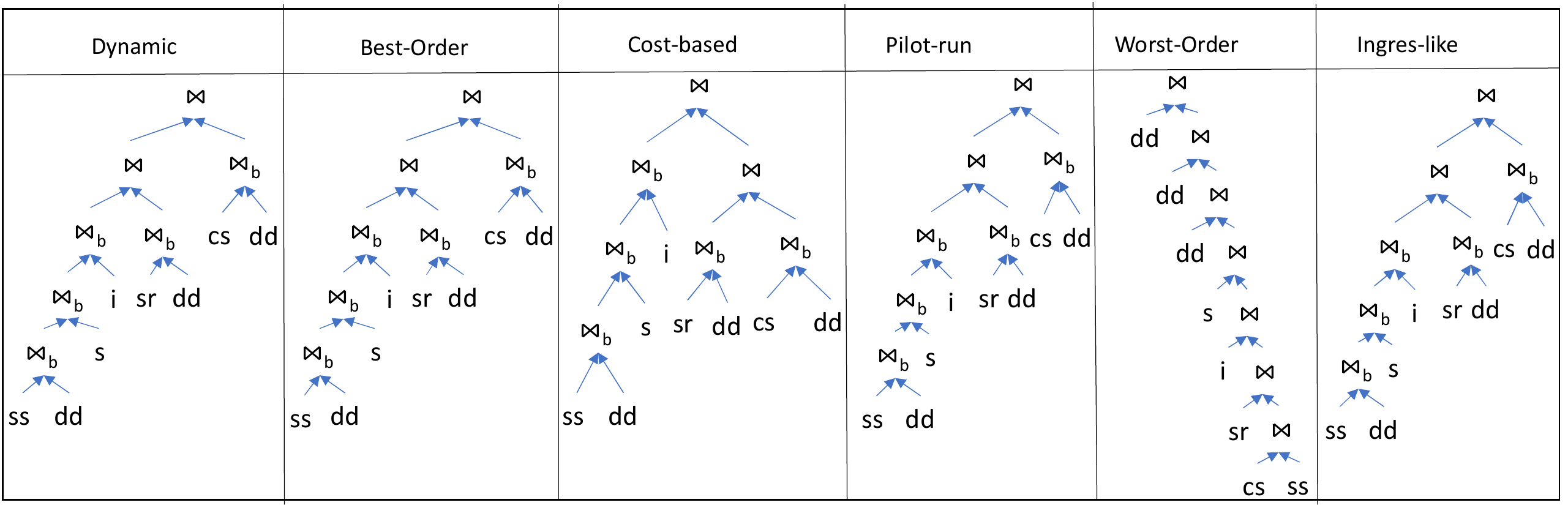}
\caption{Plans Generated for Query 17, Figure \ref{exp}, 100GB.}
\end{figure*}

\begin{figure*}[h!]
\includegraphics[width=\textwidth]{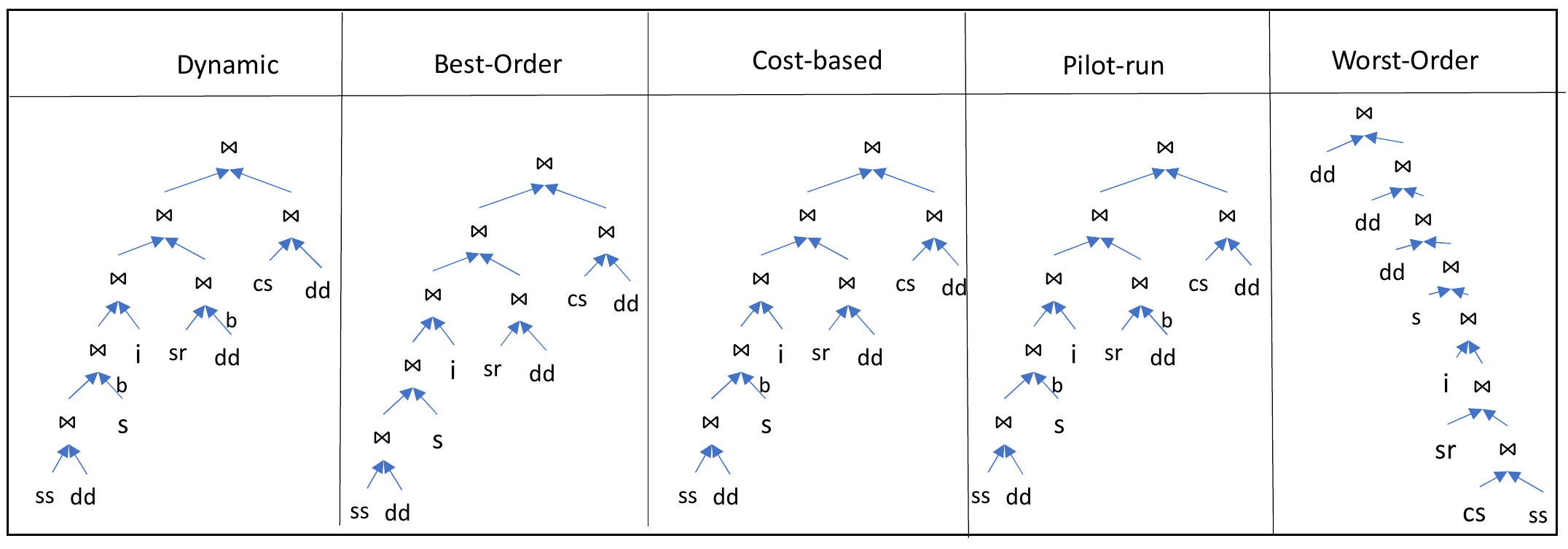}
\caption{Plans Generated for Query 17, Figure \ref{exp}, 1000GB.}
\end{figure*}




 \begin{figure*}[!tbp]
  \centering
  \begin{subfigure}[b]{0.48\textwidth}
    \includegraphics[width=\textwidth, height=4cm]{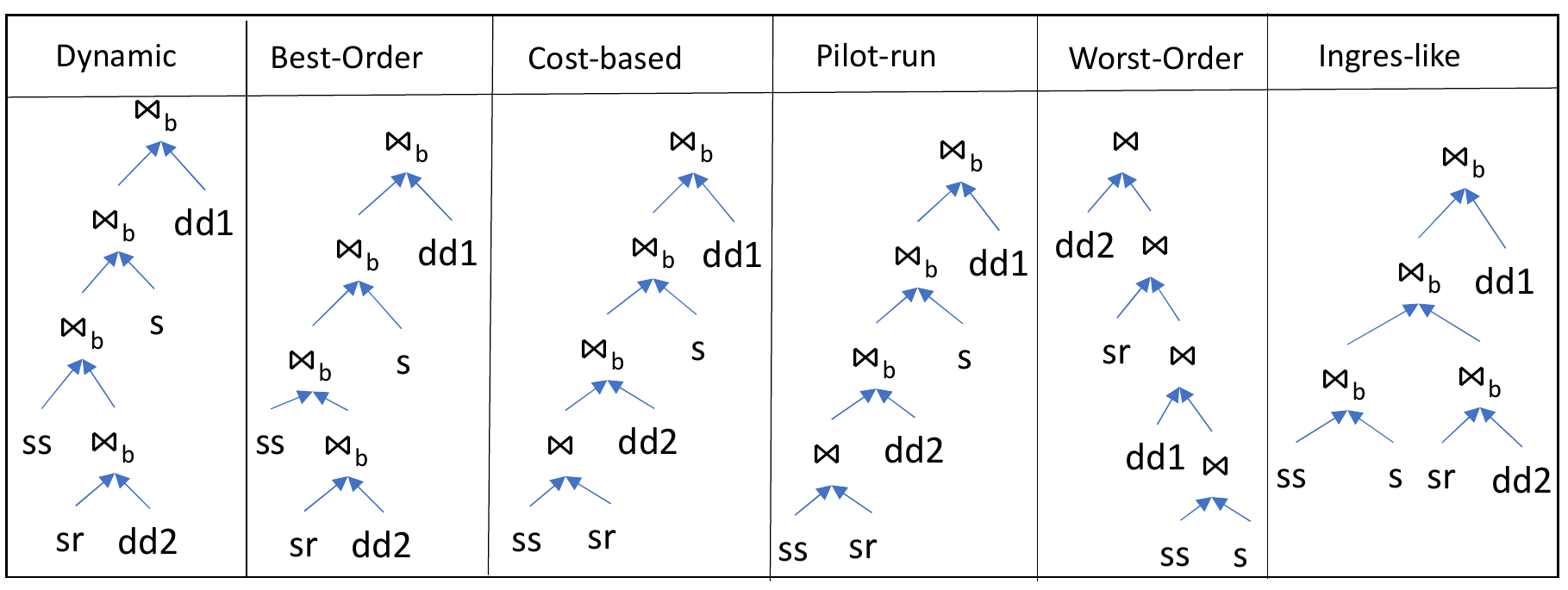}
    \caption{Scale Factor 10}
    \vspace{1em}
  \end{subfigure}
  \hfill
  \begin{subfigure}[b]{0.48\textwidth}
    \includegraphics[width=\textwidth, height=4cm]{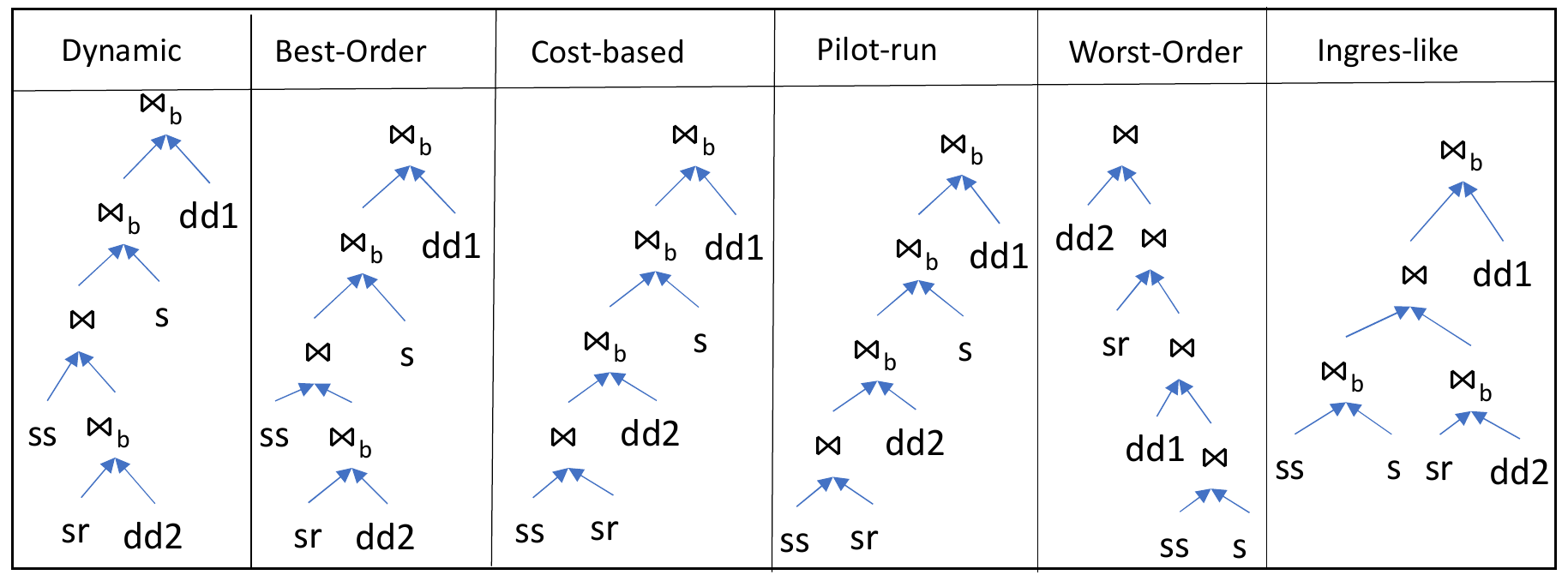}
    \caption{Scale Factor 100}
    \vspace{1em}
  \end{subfigure}
  \begin{subfigure}[b]{0.5\textwidth}
    \includegraphics[width=\textwidth, height=4cm]{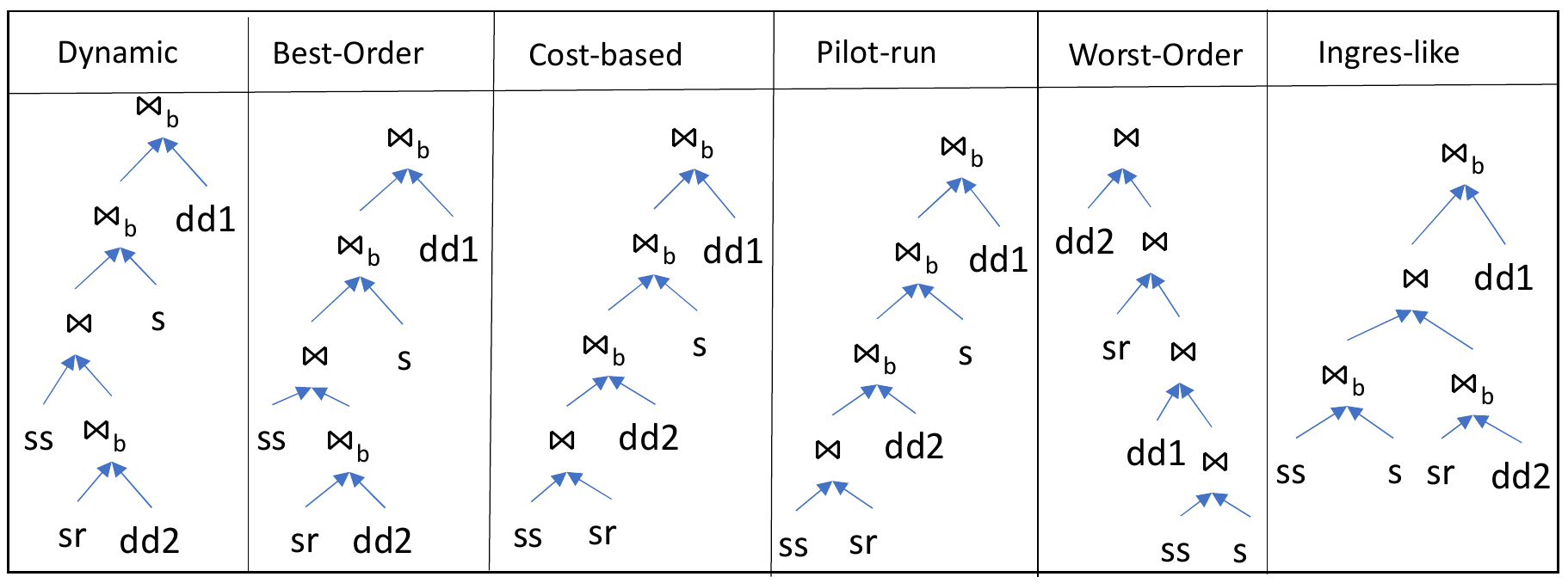}
    \caption{Scale Factor 1000}
  \end{subfigure}
  \caption{Plans Generated for Query 50, Figure \ref{exp}}
\end{figure*}

\begin{figure*}[!tbp]
  \begin{subfigure}[b]{0.48\textwidth}
    \includegraphics[width=\textwidth, height=4cm]{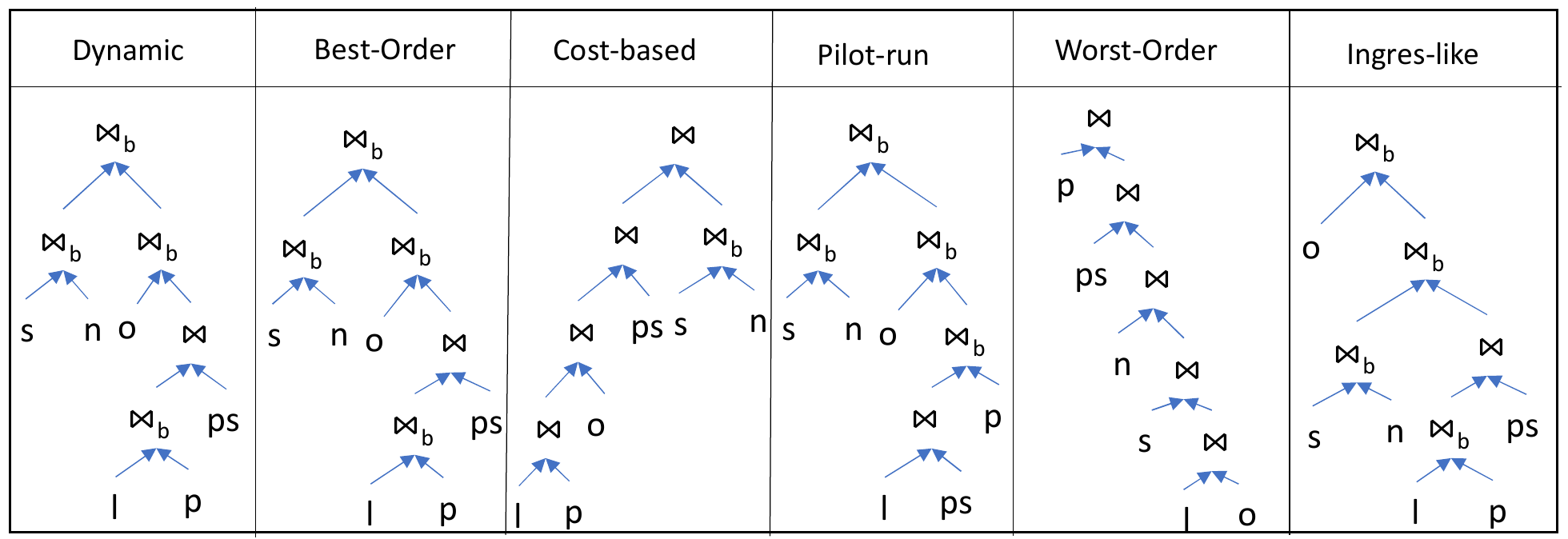}
    \caption{Scale Factor 10}
    \vspace{1em}
  \end{subfigure}
  \hfill
  \begin{subfigure}[b]{0.48\textwidth}
    \includegraphics[width=\textwidth, height=4cm]{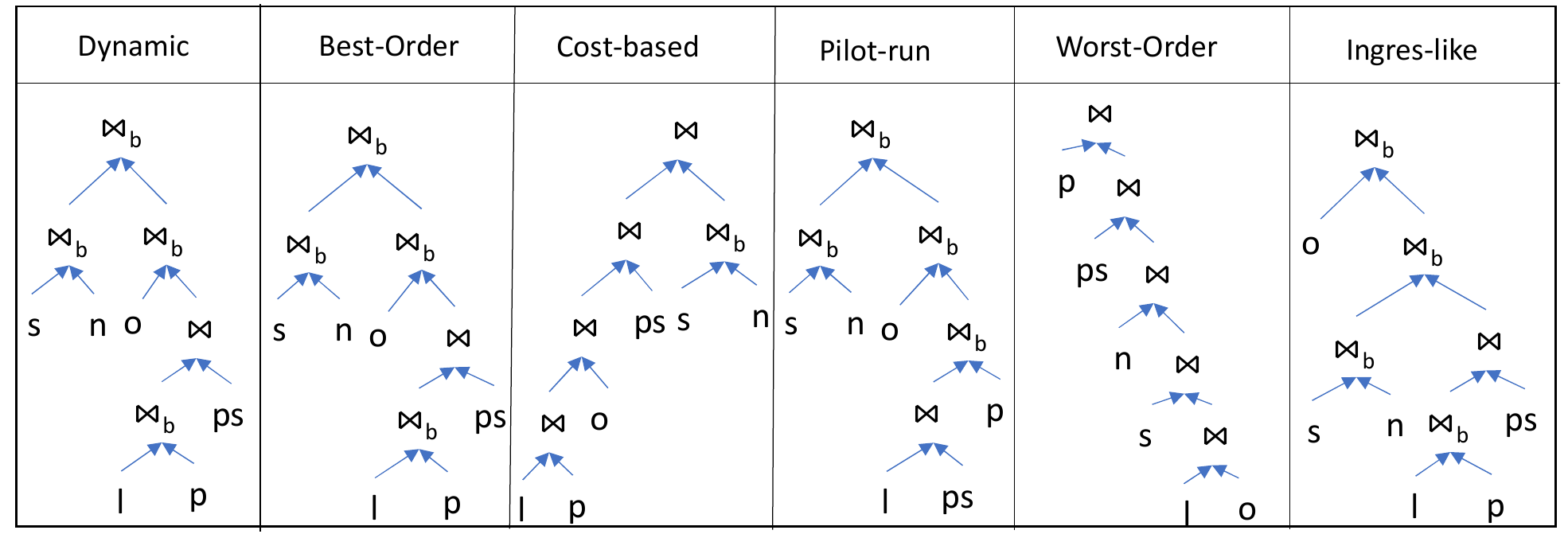}
    \caption{Scale Factor 100}
    \vspace{1em}
  \end{subfigure}
  \begin{subfigure}[b]{0.5\textwidth}
    \includegraphics[width=\textwidth, height=4cm]{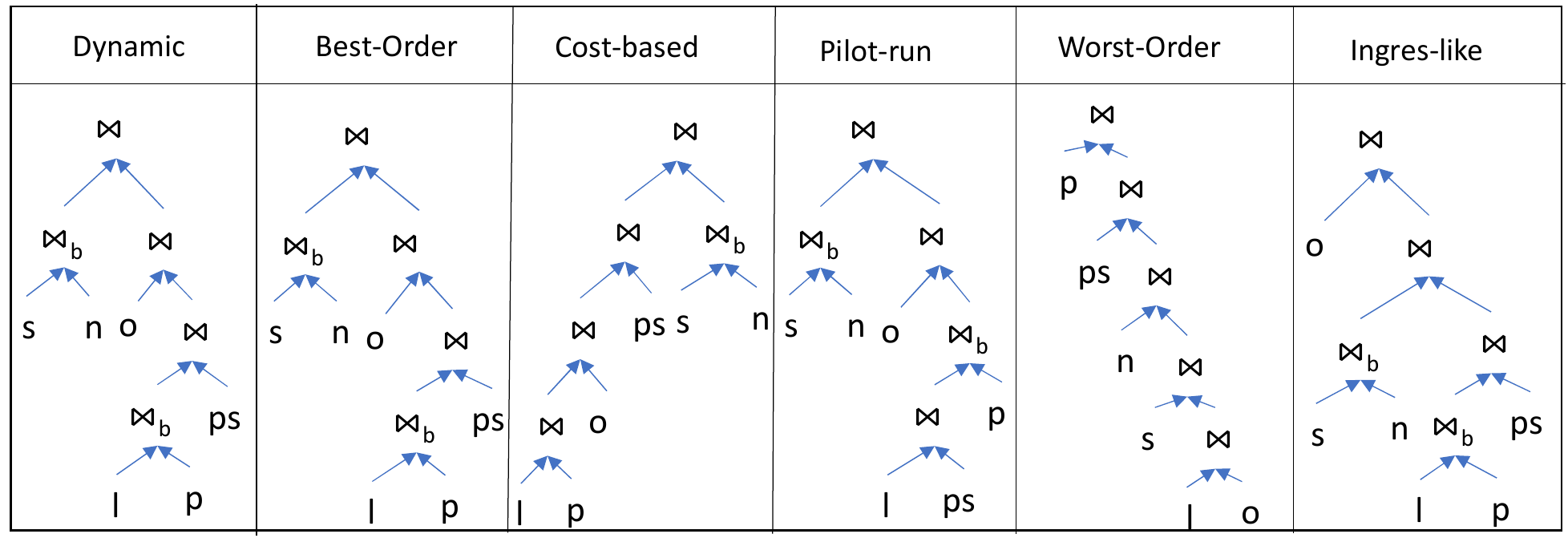}
    \caption{Scale Factor 1000}
  \end{subfigure}
  \caption{Plans Generated for Query 9, Figure \ref{exp}}
\end{figure*}


\begin{figure*}[h!]
\includegraphics[width=\textwidth]{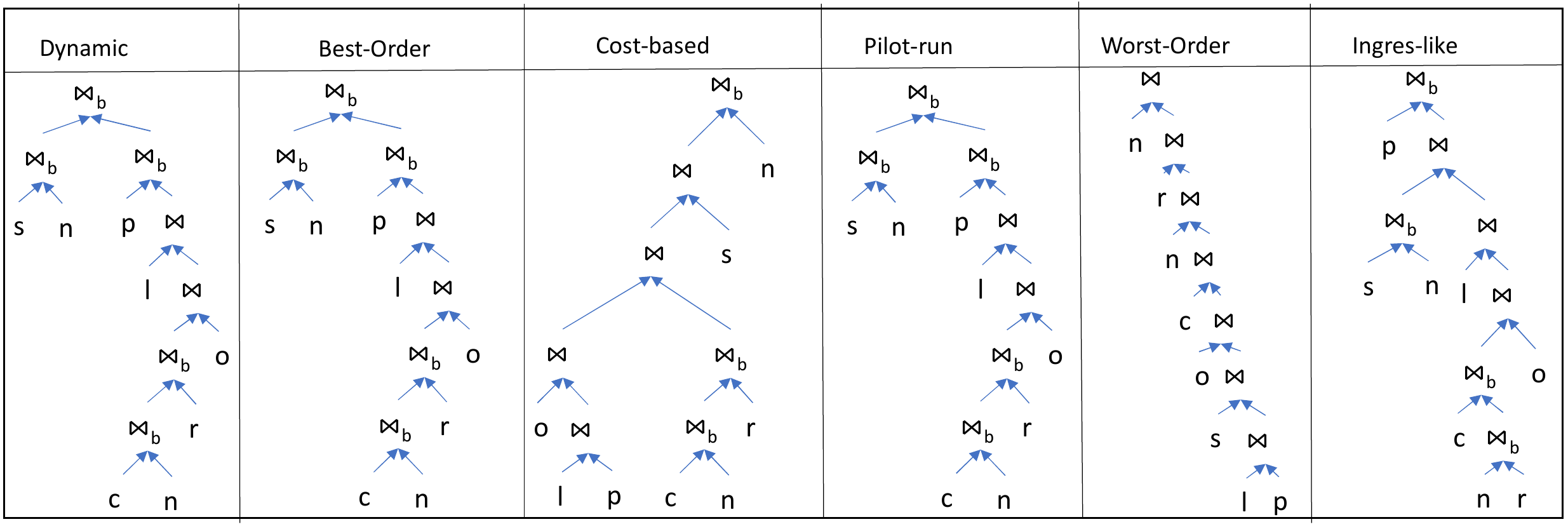}
\caption{Plans Generated for Query 8, Figure \ref{exp}, 10GB.}
\end{figure*}

\begin{figure*}[h!]
\includegraphics[width=\textwidth]{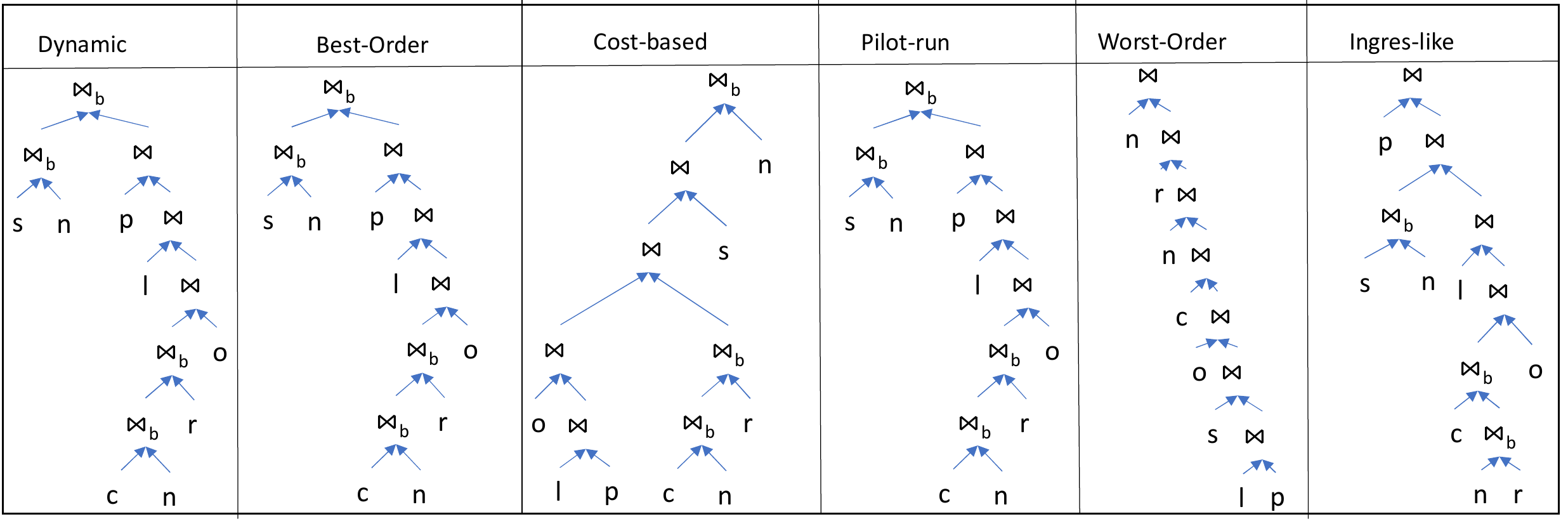}
\caption{Plans Generated for Query 8, Figure \ref{exp}, 100GB.}
\end{figure*}

\begin{figure*}[h!]
\includegraphics[width=\textwidth]{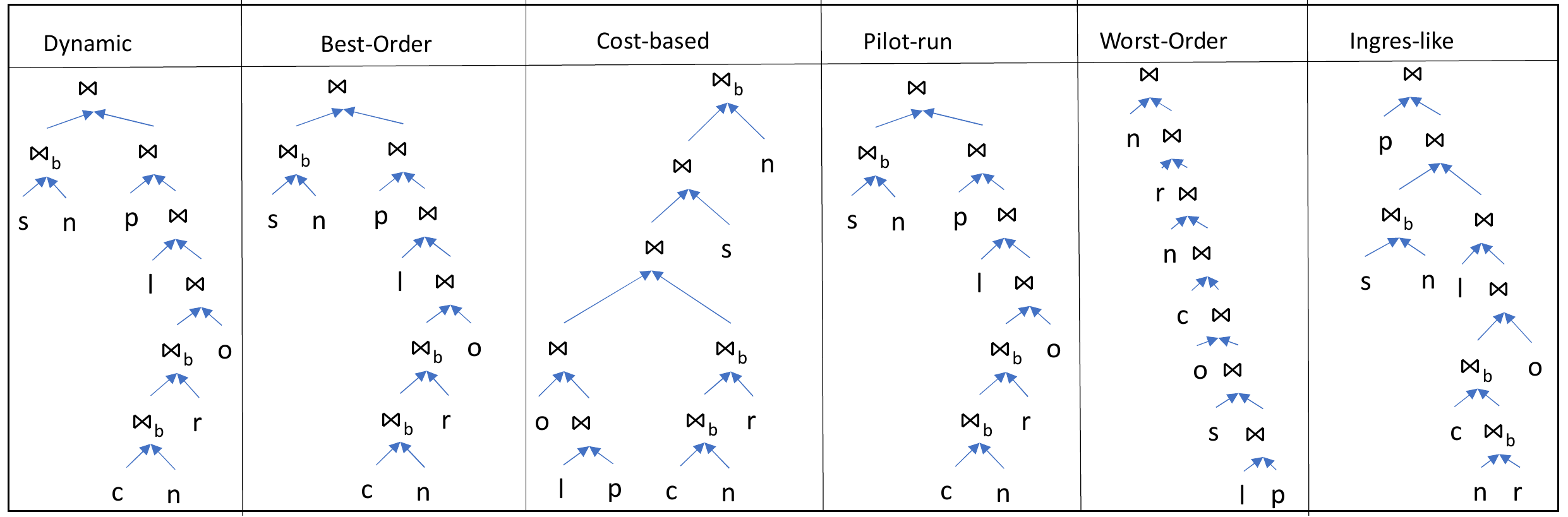}
\caption{Plans Generated for Query 8, Figure \ref{exp}, 1000GB.}
\end{figure*}





\begin{figure*}[h!]
\includegraphics[width=\textwidth]{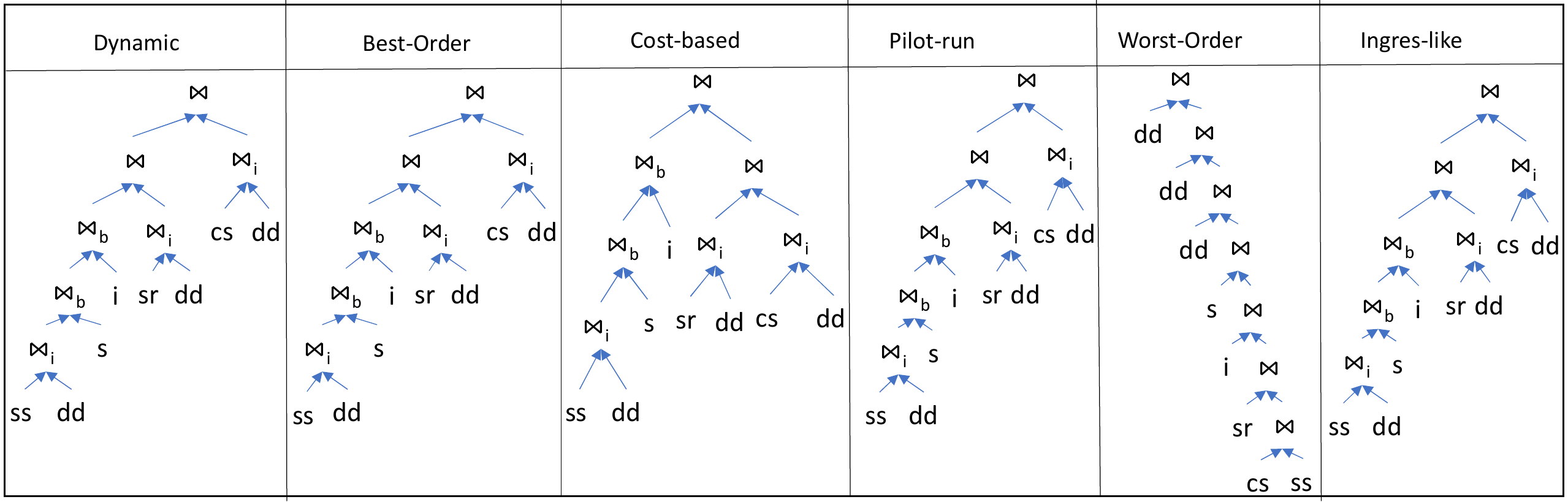}
\caption{Plans Generated for Query 17, Figure \ref{udfs}, 10GB.}
\end{figure*}

\begin{figure*}[h!]
\includegraphics[width=\textwidth]{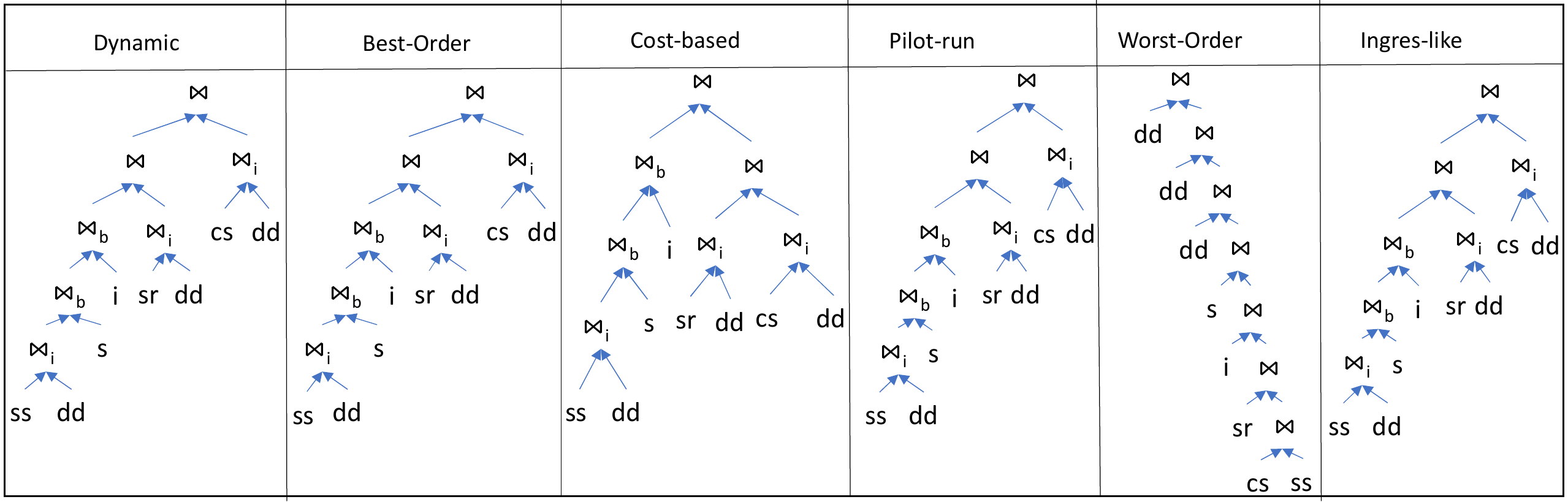}
\caption{Plans Generated for Query 17, Figure \ref{udfs}, 100GB.}
\end{figure*}

\begin{figure*}[h!]
\includegraphics[width=\textwidth]{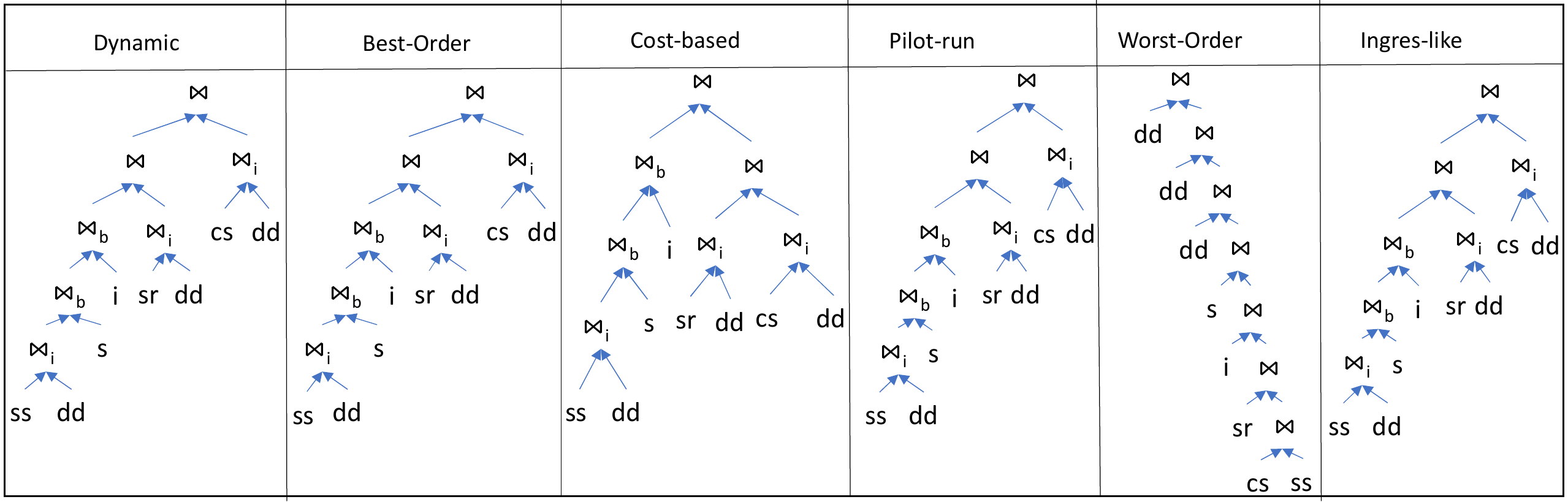}
\caption{Plans Generated for Query 17, Figure \ref{udfs}, 1000GB.}
\end{figure*}




 \begin{figure*}[!tbp]
  \centering
  \begin{subfigure}[b]{0.48\textwidth}
    \includegraphics[width=\textwidth, height=4cm]{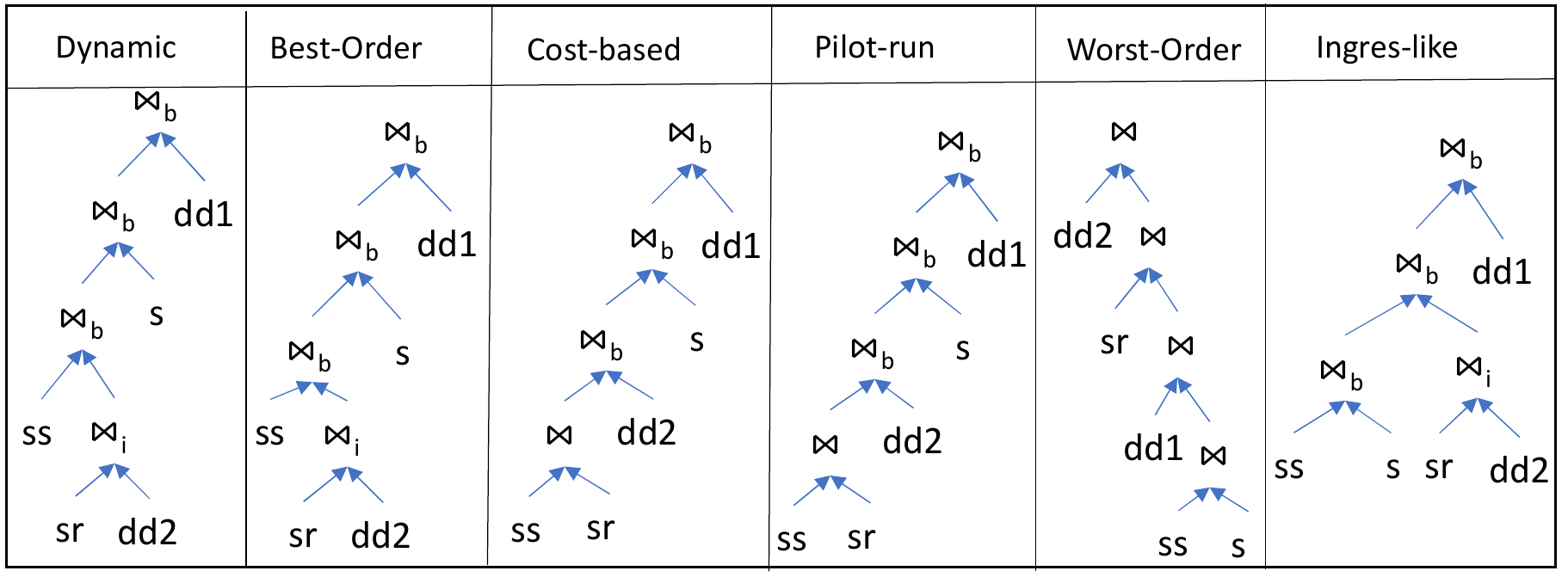}
    \caption{Scale Factor 10}
    \vspace{1em}
  \end{subfigure}
  \hfill
  \begin{subfigure}[b]{0.48\textwidth}
    \includegraphics[width=\textwidth, height=4cm]{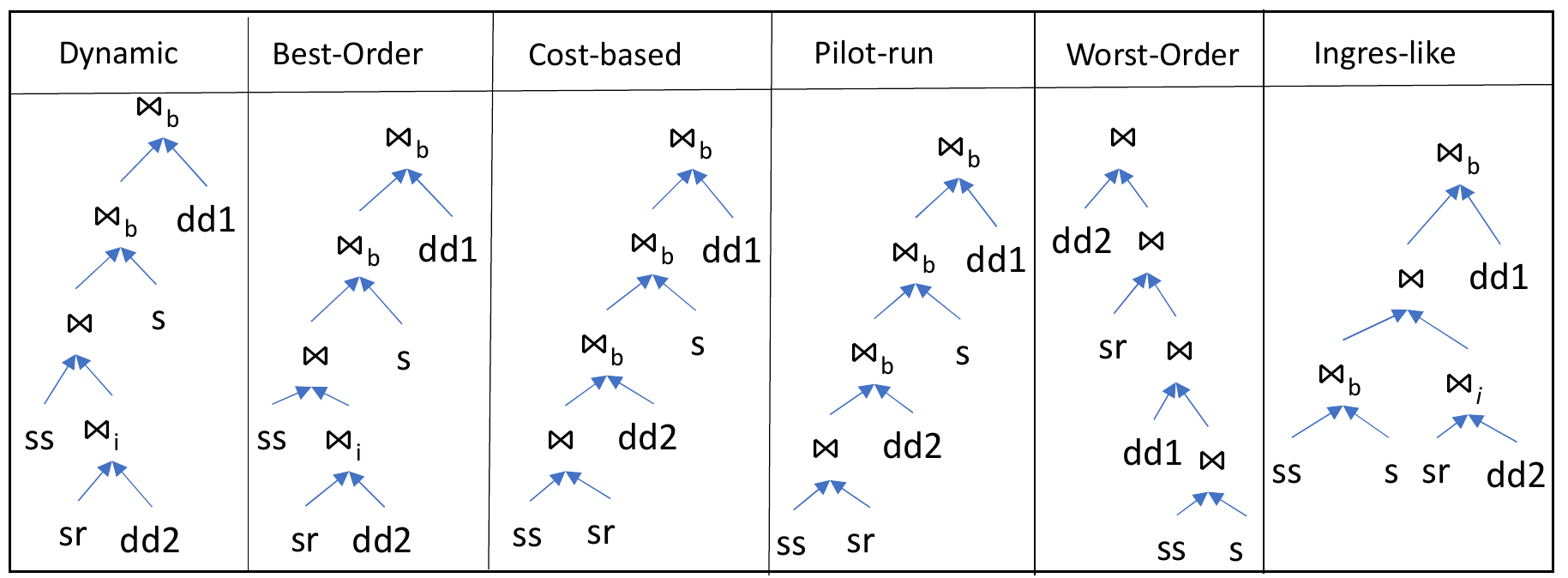}
    \caption{Scale Factor 100}
    \vspace{1em}
  \end{subfigure}
  \begin{subfigure}[b]{0.5\textwidth}
    \includegraphics[width=\textwidth, height=4cm]{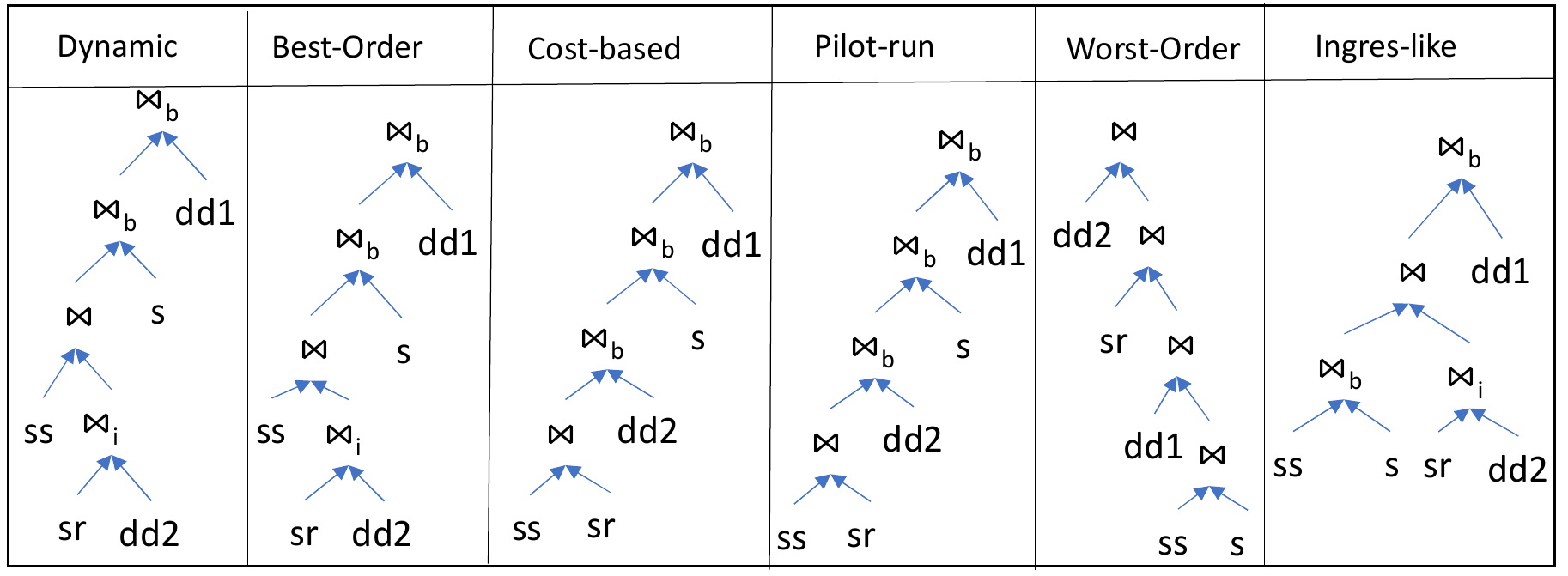}
    \caption{Scale Factor 1000}
  \end{subfigure}
  \caption{Plans Generated for Query 50, Figure \ref{udfs}}
\end{figure*}

\begin{figure*}[!tbp]
  \begin{subfigure}[b]{0.48\textwidth}
    \includegraphics[width=\textwidth, height=4cm]{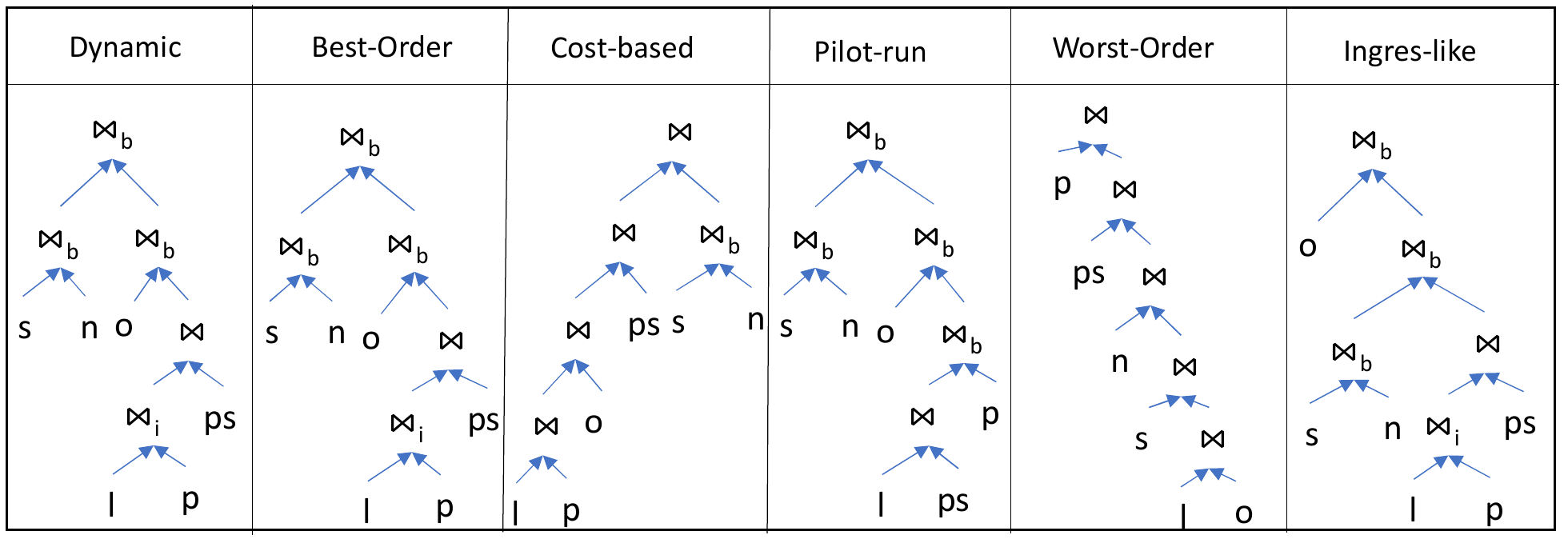}
    \caption{Scale Factor 10}
    \vspace{1em}
  \end{subfigure}
  \hfill
  \begin{subfigure}[b]{0.48\textwidth}
    \includegraphics[width=\textwidth, height=4cm]{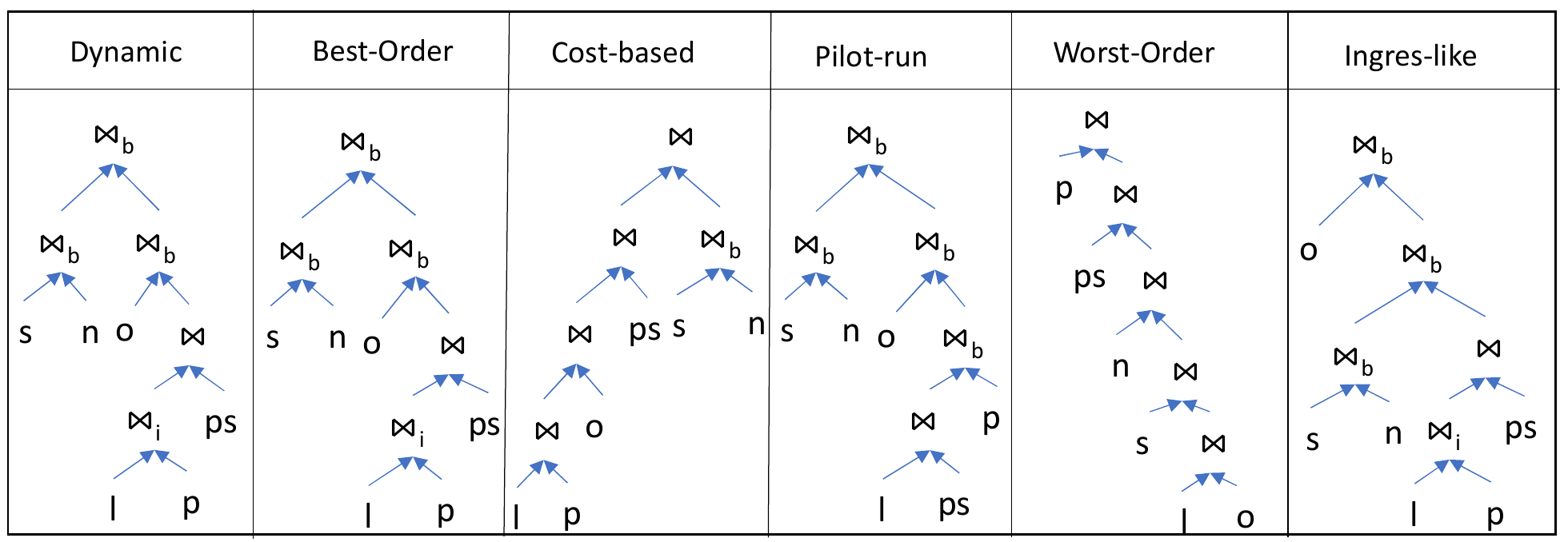}
    \caption{Scale Factor 100}
    \vspace{1em}
  \end{subfigure}
  \begin{subfigure}[b]{0.5\textwidth}
    \includegraphics[width=\textwidth, height=4cm]{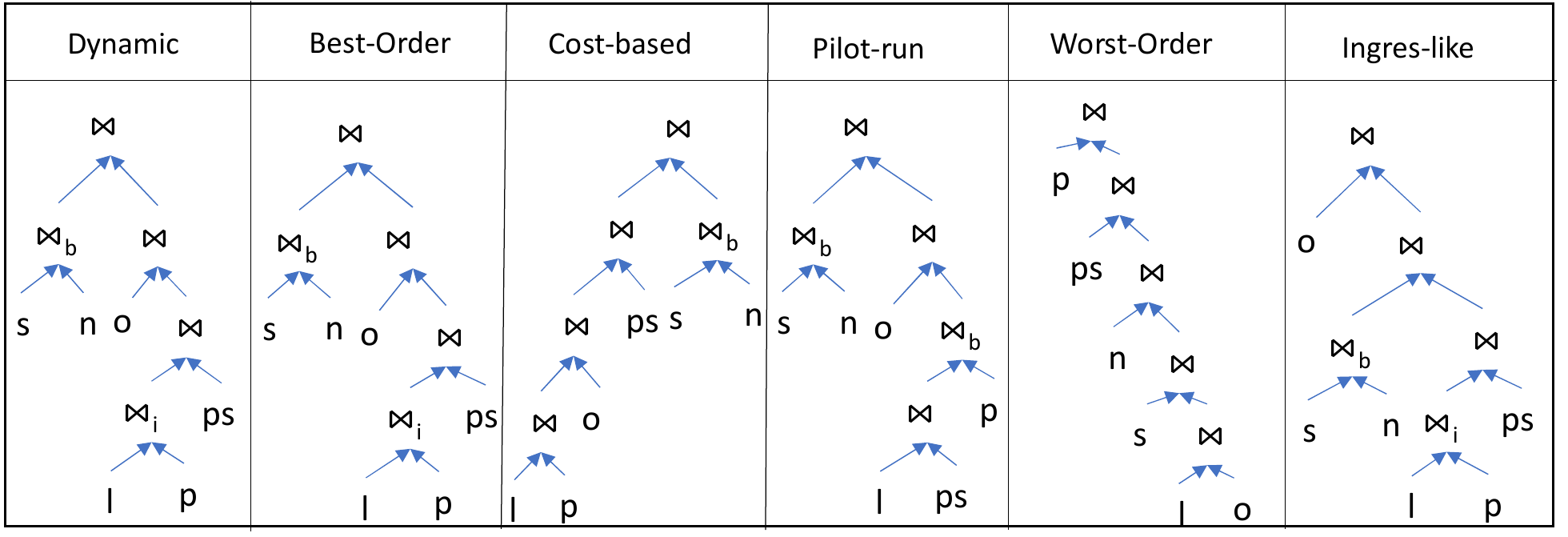}
    \caption{Scale Factor 1000}
  \end{subfigure}
  \caption{Plans Generated for Query 9, Figure \ref{udfs}}
\end{figure*}

\clearpage

\end{document}